\pdfoutput=1
\documentclass[12pt]{article}

\usepackage[T1]{fontenc}
\usepackage{lmodern}
\usepackage{microtype}
\usepackage[a4paper,margin=25mm,headheight=15pt]{geometry}
\usepackage{setspace}
\usepackage{amsmath,amssymb,bm,mathtools}
\usepackage{graphicx}
\usepackage{booktabs,tabularx,longtable,multirow,array,threeparttable}
\usepackage{xcolor}
\usepackage{subcaption}
\usepackage{enumitem}
\usepackage{placeins}
\usepackage{float}
\usepackage{pdflscape}
\usepackage{fancyhdr}
\usepackage[numbers,sort&compress]{natbib}
\usepackage{hyperref}
\usepackage[nameinlink,capitalise,noabbrev]{cleveref}

\definecolor{DeepNavy}{HTML}{17365D}
\definecolor{SignalTeal}{HTML}{007C83}
\definecolor{WarmOrange}{HTML}{D8742F}
\definecolor{SoftGray}{HTML}{5B6573}
\hypersetup{
  colorlinks=true,
  linkcolor=DeepNavy,
  citecolor=SignalTeal,
  urlcolor=black,
  pdfauthor={Ehsan Roohi},
  pdftitle={Geometry-native machine learning reconstruction of DSMC moment fields with support monitoring}
}

\graphicspath{{figures_journal/}}
\newcolumntype{Y}{>{\raggedright\arraybackslash}X}
\newcommand{\nrmse}{\operatorname{NRMSE}}
\newcommand{\Raw}[1]{\ensuremath{\mathrm{Raw}(#1)}}
\newcommand{\doi}[1]{\href{https://doi.org/#1}{\nolinkurl{doi:#1}}}
\newenvironment{journalfigurepage}
  {\clearpage\begin{landscape}\thispagestyle{plain}\begin{figure}[p]\centering\captionsetup{font=normalsize}}
  {\end{figure}\end{landscape}\clearpage}
\newcommand{\journalgraphic}[1]{%
  \makebox[\linewidth][c]{\includegraphics[width=1.15\linewidth,height=0.815\textheight,keepaspectratio]{#1}}}

\title{\vspace{-1.6cm}\textbf{Geometry-native machine learning reconstruction of DSMC moment fields with support monitoring}}
\author{Ehsan Roohi\\[-0.1em]
\small Department of Mechanical and Industrial Engineering\\[-0.1em]
\small University of Massachusetts Amherst, 160 Governors Drive, Amherst, MA 01003, USA\\[-0.1em]
\small \href{mailto:roohie@umass.edu}{roohie@umass.edu}\quad ORCID: \href{https://orcid.org/0000-0001-5739-3210}{0000-0001-5739-3210}}
\date{}

\begin{document}
\maketitle

\begin{abstract}\small
Direct simulation Monte Carlo (DSMC) resolves rarefied-gas dynamics without a constitutive closure, but finite-sample estimates of macroscopic moments converge at markedly different rates.  We develop a non-intrusive, geometry-native machine learning reconstruction of the retained two-dimensional moment hierarchy: number density, two velocity components, translational temperature, three pressure-tensor components, and two heat-flux components.  From three sampling blocks, the estimator corrects a structured prior learned from development data with a bounded term computed from the current observation, while preserving additive-moment consistency and the measured zero-frequency content.  In cavity development tests, the final observation-conditioned estimator, whose prior is a trained MambaIR restoration network, reduces transverse-heat-flux error to 0.658 and 0.672 times that of a ten-block direct average at two rarefied conditions.  For a hypersonic cylinder, a cylinder-centred estimator is fixed before evaluation on six new observation/reference pairs.  It improves both global transverse heat flux and near-wall normal heat flux in every pair; the ratios of arithmetic-mean normalised root-mean-square errors (NRMSEs) are 0.846 and 0.793, and the Holm-adjusted one-sided exact probabilities are 0.03125.
A reference-independent support monitor is calibrated from eight Mach-8 and Mach-10 development trajectories.  It accepts all six held-out Mach-10 observations and classifies all four subsequently generated Mach-12 observations as outside support before their references are accessed.  Post-decision scoring shows that the abstained Mach-12 estimate still lowers three-block raw error by 5.8\% across the nine-field geometric mean; an estimate can be numerically useful outside its validated domain.  With separate same-condition Mach-12 calibration, all nine fields are reconstructed at approximately one-half of the ten-block NRMSE.  The outcome is a reduced-sampling reconstruction with an explicit domain-of-validity decision; geometry dependence, finite-reference uncertainty, and the separate statistics of wall-collision heat flux remain its stated limitations.
\end{abstract}


\section{Introduction}
\label{sec:introduction}

Direct simulation Monte Carlo (DSMC) is the standard particle method for rarefied and nonequilibrium gas dynamics \citep{Bird1994DSMC}.  Its macroscopic fields are statistical estimates, and their sampling errors differ substantially with moment order \citep{HadjiconstantinouEtAl2003StatisticalError}.  Number density and bulk velocity often converge long before stress and heat flux.  Heat flux is a signed third-order central moment: large molecular contributions cancel to produce a comparatively weak mean, so a sampling budget that is adequate for density or temperature may remain unusably noisy for heat-transfer analysis.  The computational question addressed in this paper is how much of a physically meaningful DSMC moment hierarchy can be recovered from a short sampling window by a machine-learning estimator and, no less important, how a deployed estimator can decide, without access to a reference solution, whether its reconstruction should be trusted at the current operating condition.

A mature body of work reduces DSMC noise by modifying the particle calculation itself.  Control variates and deviational formulations simulate the departure from an equilibrium or reference state \citep{BakerHadjiconstantinou2005VarianceReduction,HomolleHadjiconstantinou2007JCP,HomolleHadjiconstantinou2007PoF}; importance weights, relaxation-time and variable-hard-sphere formulations, and multi-effect or binary-mixture extensions broaden the accessible regimes \citep{RadtkeHadjiconstantinou2009VarianceReducedRTA,AlMohssenHadjiconstantinou2010ImportanceWeights,HadjiconstantinouEtAl2010HeatTransferVR,RadtkeEtAl2011LowNoiseVHS,Szalmas2012BinaryVarianceReduced,SadrHadjiconstantinou2023MEVRDSMC,SadrHadjiconstantinou2023FokkerPlanckVR}.  Most recently, denoising multiscale particle methods embed low-noise collective information directly in the simulator particles \citep{YangEtAl2025DenoisingParticle,YangZhang2026MSMC}.  These methods can be highly effective, but they require a modified solver, a reference distribution, or online model integration.  The problem considered here is complementary: a production DSMC solver has already written additive molecular moments, and the objective is to reconstruct the complete retained hierarchy non-intrusively, after sampling, from a short window.

Non-intrusive filtering of particle fields also has direct precedent in the literature: \citet{ZimonEtAl2016NoiseReduction} systematically evaluated proper-orthogonal-decomposition and wavelet-type noise-reduction algorithms for DSMC and molecular-dynamics fields in multiscale applications, building on the classical estimation toolbox of low-rank projection, Wiener filtering, and transform-domain shrinkage \citep{BerkoozEtAl1993POD,Wiener1949Filtering,AhmedNatarajanRao1974DCT,ChungEtAl2011OptimalSpectralFilters}.  That generation of methods left open three questions that motivate the present work.  Generic filters use no information beyond the current realisation, although repeatable flow structure across operating conditions and random seeds is an exploitable prior.  A filter tuned to look smooth can also silently corrupt physically meaningful low-frequency content, the amplitude and spatial mean of a weak signed moment, which is precisely the content that matters for cancellation-dominated heat flux.  Most important for deployment, none of these estimators reports \emph{when it should not be used}: a prior-based reconstruction that is accurate at its development conditions may remain visually plausible, yet biased, after the operating condition shifts.

Machine-learning reconstruction of fluid fields from sparse or noisy data is now an established research direction \citep{FukamiEtAl2023SRSurvey}.  Super-resolution and self-supervised denoising have been demonstrated for turbulence \citep{FukamiEtAl2019SuperResolution,FukamiEtAl2021SpatiotemporalSR,KimEtAl2021UnsupervisedTurbulenceSR,YuEtAl2024SelfSupervisedFlowDenoising}, sparse-sensor reconstruction via sparse representation and physics-informed networks \citep{CallahamEtAl2019SparseFlowReconstruction,GaoEtAl2021PhysicsInformedFlowSR,FathiEtAl2020PhysicsInformed4DFlow,ShuEtAl2023PhysicsInformedDiffusion,MorenoSotoEtAl2024CompleteFlow,RowanEtAl2025ExplicitConstraintForce,CestolaEtAl2026FlowAwarePINN}, and hard measurement constraints have proved decisive under noise: \citet{MoMagri2025PhysicsConstrainedFlow} show that enforcing the \emph{mean} of the prediction at sensor locations is the robust choice for noisy data, and extend the approach to three-dimensional reconstruction \citep{MoMagri2026WeightSharing3D}.  The data-consistency operation used here belongs to the same family: the observed zero-frequency content is preserved exactly while a bounded transform-domain residual corrects the prior, a construction that inherits its rationale from regularised and learned inverse problems \citep{ArridgeEtAl2019DataDrivenInverse,AggarwalEtAl2019MoDL}.  Self-supervised training without clean targets follows the noise-to-noise principle \citep{LehtinenEtAl2018Noise2Noise,BatsonRoyer2019Noise2Self}, and the structured priors are drawn from modern restoration backbones \citep{ChenEtAl2022NAFNet,GuoEtAl2024MambaIR}.  Particle moment fields, however, differ from images and velocimetry planes in ways that this literature does not address: the noise is strongly heteroscedastic, the reported quantities share nonlinear centralisation operations acting on common additive accumulators, the natural representation changes with geometry, and physics-informed residuals face an identifiability barrier discussed next.  Physics-informed and operator-learning methods \citep{RaissiEtAl2019PINN,KarniadakisEtAl2021PIML,LuEtAl2021DeepONet,LouEtAl2021BoltzmannBGKPINN}, including recent kinetic surrogates and learned collision models from our group \citep{Roohi2026NeuralCollision,RoohiShojaSani2026DSMCSurrogate,Roohi2026RarefiedNeuralNetworks,RoohiMahdavi2026MicroStepDeepONet,Roohi2026GPUFokkerPlanckJCP}, accelerate the forward kinetic problem; the present work instead treats the statistical inverse problem attached to an unmodified solver, and uses conservation identities diagnostically rather than as training losses, wary of the optimisation pathologies such residuals can introduce \citep{WangEtAl2022PINNFailure}.

A smooth heat-flux field can carry the wrong amplitude or spatial mean, and an exact energy-balance residual cannot uniquely repair the error: in two dimensions, energy conservation constrains the divergence of the heat-flux vector but is insensitive to its solenoidal part, including a constant component offset.  ``Adding physics'' through a conservation penalty therefore does not identify the difficult component; the missing information must come from the current observation.  The estimator built here pairs a structured prior learned from development trajectories with a bounded residual formed from the present DSMC observation and preserves the observed zero-frequency content exactly.  All central moments are formed only after additive accumulators have been combined, which prevents the systematic bias produced by centralising individual sampling blocks before averaging.

Reduced error at a tested condition does not establish unrestricted deployment.  We accompany the reconstruction with a reference-independent support monitor: an empirical-Bayes observation gain is computed for nine field components in three geometric zones, and a fixed development-envelope threshold is calibrated by a leave-one-unit-out maximum-score rule on Mach-8 and Mach-10 trajectories.  The construction parallels split-conformal calibration \citep{Vovk2005ALRW,ShaferVovk2008Tutorial,AngelopoulosBates2023CP}, applied to a familywise gain score rather than a pointwise nonconformity score, and, to our knowledge, has no precedent as an accept/abstain rule for reconstruction attached to a particle kinetic solver.  Specificity is tested on held-out Mach-10 observations, and shift detection is tested on subsequently generated Mach-12 observations whose support decisions are locked before any paired reference is accessed.  This separates two questions that are often conflated: whether a reconstruction is numerically useful, and whether it belongs to the regime in which its error reduction has been validated.

The method is developed on a rarefied lid-driven cavity and reformulated for hypersonic argon flow over a circular cylinder.  The cavity stage uses the MambaIR state-space restoration model \citep{GuoEtAl2024MambaIR} as a structural prior and a discrete cosine transform to separate reliable low modes from sampling-dominated content; the cylinder stage uses native cell areas, cylinder-centred normal and tangential heat flux, and a bounded two-component spectral transfer.  Nine two-dimensional fields are retained: number density, the two velocity components, translational temperature, three pressure-tensor components, and the two heat-flux components.  A direct estimate constructed from $B$ additive sampling blocks is denoted $\mathrm{Raw}(B)$; the principal low-budget estimate is \Raw{3} and is compared with \Raw{10}.  The final cavity estimator reduces transverse-heat-flux error to 0.658 and 0.672 times the ten-block error at the two analysed conditions.  The fixed cylinder estimator then improves global transverse heat flux and near-wall normal heat flux in all six independent Mach-10 pairs, with arithmetic-mean normalised root-mean-square error (NRMSE) ratios of 0.846 and 0.793 and Holm-adjusted one-sided exact probabilities of 0.03125.  The support monitor, finally, accepts all six held-out Mach-10 observations and abstains on all four fresh Mach-12 observations before reference evaluation; a separately labelled same-condition Mach-12 analysis reconstructs all nine retained fields at approximately one-half of the ten-block error.

Relative to the intrusive variance-reduction line \citep{HomolleHadjiconstantinou2007JCP,SadrHadjiconstantinou2023MEVRDSMC,YangEtAl2025DenoisingParticle}, to non-intrusive filter evaluations \citep{ZimonEtAl2016NoiseReduction}, and to learned flow reconstruction \citep{FukamiEtAl2023SRSurvey,MoMagri2025PhysicsConstrainedFlow}, the paper makes five contributions:
\begin{enumerate}
  \item a non-intrusive, geometry-native machine learning reconstruction of the complete retained two-dimensional moment hierarchy from three sampling blocks, built on additive-first centralisation that is exact by construction;
  \item a data-consistent fusion rule in which a frozen structured prior is corrected by a bounded transform-domain residual while the measured zero-frequency content is preserved exactly, together with an identifiability analysis showing why energy conservation alone cannot recover a heat-flux component;
  \item a reference-independent domain-of-validity monitor with development-only calibration, validated by a held-out specificity test (six of six supported observations accepted) and a prospective condition-shift test (four of four fresh Mach-12 observations abstained before reference access);
  \item a prospective, seed-level statistical design, frozen estimators, disjoint observation/comparator/reference partitions, and exact paired sign tests, that separates development, attribution, and confirmation evidence and quantifies how much of each gain is supplied by the prior versus the current observation;
  \item boundary-specific statistical findings of independent practical value, including the measurement that the native wall-collision heat-flux tally retains an effective sample size an order of magnitude smaller than neighbouring volume fields.
\end{enumerate}

The remainder of the paper defines the kinetic observables and estimator, describes the numerical and validation design, presents the cavity and cylinder results, and discusses bias, support, finite-reference uncertainty, and the distinction between cell-centred and wall-collision heat flux.

\section{Kinetic observables and statistical formulation}
\label{sec:formulation}

\subsection{Moment hierarchy and additive aggregation}

Let $f(\boldsymbol x,\boldsymbol\xi,t)$ be the one-particle velocity distribution, where $\boldsymbol x\in\Omega\subset\mathbb R^2$, $\boldsymbol\xi\in\mathbb R^3$, and $t$ denote position in the two-dimensional computational domain $\Omega$, three-dimensional molecular velocity, and time.  Thus the simulations are spatially two-dimensional with three velocity components.  For a monatomic gas of molecular mass $m_{\mathrm{mol}}$, define the bulk velocity $\boldsymbol u$, peculiar velocity $\boldsymbol c=\boldsymbol\xi-\boldsymbol u$, pressure tensor $\boldsymbol P$, heat-flux vector $\boldsymbol q$, and translational temperature $T$ by
\begin{align}
 n &= \int_{\mathbb R^3} f\,\mathrm d\boldsymbol\xi,
 & n\boldsymbol u &= \int_{\mathbb R^3}\boldsymbol\xi f\,\mathrm d\boldsymbol\xi, \\
 \boldsymbol P &= m_{\mathrm{mol}}\int_{\mathbb R^3}\boldsymbol c\boldsymbol c f\,\mathrm d\boldsymbol\xi,
 & \boldsymbol q &= \frac{m_{\mathrm{mol}}}{2}\int_{\mathbb R^3}|\boldsymbol c|^2\boldsymbol c f\,\mathrm d\boldsymbol\xi, \\
 T &= \frac{\operatorname{tr}(\boldsymbol P)}{3n k_B},
\label{eq:kinetic-moments-r14}
\end{align}
where $n$ is number density, $k_B$ is the Boltzmann constant, $\operatorname{tr}$ denotes the tensor trace, and juxtaposition of two vectors denotes their dyadic product.  No Fourier-law or other continuum closure is invoked.  The retained two-dimensional field vector is
\begin{equation}
 \boldsymbol m=(n,u,v,T,P_{xx},P_{xy},P_{yy},q_x,q_y)^{\mathsf T},
 \label{eq:field-vector-r14}
\end{equation}
where $u$ and $v$ are the Cartesian $x$- and $y$-components of $\boldsymbol u$ and the superscript $\mathsf T$ denotes transpose.

The solver stores additive rather than central moments.  For simulator particles indexed by $p$, with statistical weights $w_p$, define
\begin{align}
 C_0 &= \sum_p w_p, & C_i &= \sum_p w_p \xi_{p,i},
 & C_{ij} &= \sum_p w_p \xi_{p,i}\xi_{p,j}, \\
 E_2 &= \sum_p w_p|\boldsymbol\xi_p|^2,
 & F_i &= \sum_p w_p|\boldsymbol\xi_p|^2\xi_{p,i},
\label{eq:additive-r14}
\end{align}
where $i,j\in\{x,y,z\}$ label Cartesian components.  Here $C_0$, $C_i$, and $C_{ij}$ are the weighted zeroth-, first-, and second-order raw moments, $E_2$ is the squared-speed moment, and $F_i$ is the corresponding raw energy-flux moment.  After summing these quantities over the selected blocks, $u_i=C_i/C_0$ and the central third-order numerator is evaluated once as
\begin{equation}
 J_i=F_i-u_iE_2-2\sum_j u_jC_{ji}+2C_0u_i|\boldsymbol u|^2.
 \label{eq:central-third-r14}
\end{equation}
The physical heat flux is obtained from $J_i$ using the solver's common molecular-mass, cell-volume, and sampling-time factors.  Centralising each block separately and then averaging would insert a different block velocity into the nonlinear polynomial in \cref{eq:central-third-r14}; it is therefore not equivalent to additive-first aggregation.  All fields in this work are formed by summing the accumulators in \cref{eq:additive-r14} before any centralisation.

Heat flux is statistically demanding because the terms on the right of \cref{eq:central-third-r14} can be individually large while their signed sum is small.  Modest absolute errors in bulk velocity or raw energy flux can therefore be a large fraction of $q_x$ or $q_y$.  This cancellation explains why spatial correlation and visual smoothness are insufficient diagnostics: a coherent offset or amplitude error can dominate the normalised field error even when the principal pattern is correct.

\subsection{Heat-flux identifiability}

The exact monatomic internal-energy balance is
\begin{equation}
 \frac{3}{2}nk_B\frac{DT}{Dt}
 +\boldsymbol P:\nabla\boldsymbol u+\nabla\!\cdot\boldsymbol q=0,
 \label{eq:energy-r14}
\end{equation}
where $D/Dt=\partial_t+\boldsymbol u\cdot\nabla$ is the material derivative and $:$ denotes double contraction.  Although \cref{eq:energy-r14} is an exact kinetic moment identity, it supplies only one scalar constraint for the two in-plane heat-flux components.  For any sufficiently smooth scalar field $\psi$,
\begin{equation}
 \delta\boldsymbol q=\nabla^\perp\psi
 =\left(\frac{\partial\psi}{\partial y},-\frac{\partial\psi}{\partial x}\right),
 \qquad \nabla\!\cdot\delta\boldsymbol q=0.
 \label{eq:null-r14}
\end{equation}
Thus $\boldsymbol q$ and $\boldsymbol q+\delta\boldsymbol q$ have the same energy residual.  A constant error in one component is a simple member of this null space.  Conservation is retained as an audit of the reconstructed hierarchy, but it is the current observation, not residual minimisation, that must anchor the unidentifiable content.

\subsection{Observations, references, and error measure}

A direct field obtained from $B$ declared additive blocks is denoted $\mathrm{Raw}(B)$; for example, \Raw{3} is the average formed from three blocks.  Its abstract observation model is
\begin{equation}
 \boldsymbol m^{\mathrm{obs}}_B=\boldsymbol m^{\mathrm{true}}+\boldsymbol\varepsilon_B,
 \qquad \mathbb E[\boldsymbol\varepsilon_B]\simeq\boldsymbol 0,
 \label{eq:observation-r14}
\end{equation}
where $\boldsymbol m^{\mathrm{true}}$ is the infinite-sampling expectation, $\boldsymbol\varepsilon_B$ is the finite-sampling error, and $\mathbb E$ denotes expectation.  The covariance of $\boldsymbol\varepsilon_B$ depends on particle population, moment order, temporal correlation, and block construction.  The references used below are independent or leave-one-seed-out DSMC averages, not exact solutions.  The common independent reference makes each comparison paired and does not preferentially enter either prediction, but the resulting error is measured against a stochastic reference, not against an exact Boltzmann solution.

For a scalar field $a$, reference $a^{\mathrm{ref}}$, fluid-cell index $c$, cell area $A_c$, and evaluation mask $\mathcal M$, the area-weighted normalised root-mean-square error (NRMSE) is
\begin{equation}
 \nrmse_{A,\mathcal M}(a,a^{\mathrm{ref}})=
 \left[
 \frac{\sum_{c\in\mathcal M}A_c(a_c-a_c^{\mathrm{ref}})^2}
 {\sum_{c\in\mathcal M}A_c(a_c^{\mathrm{ref}})^2}
 \right]^{1/2}.
 \label{eq:nrmse-r14}
\end{equation}
Equal-area cavity cells reduce \cref{eq:nrmse-r14} to the usual unweighted expression.  For evaluation unit $j$, the estimator-to-comparator ratio for field $a$ is
\begin{equation}
 r_{a,j}=\frac{\nrmse(a^{\mathrm{est}}_j,a^{\mathrm{ref}}_j)}
 {\nrmse(a^{\Raw{10}}_j,a^{\mathrm{ref}}_j)}.
 \label{eq:ratio-r14}
\end{equation}
A value below one indicates lower error than the ten-block direct comparator against the same reference.  The units of inference are independent trajectories or seeds, not spatial cells.

For the cylinder co-primary endpoints, the reported aggregate ratio is the ratio of arithmetic-mean NRMSEs,
\begin{equation}
 \mathcal R_a=
 \frac{N_{\mathrm{pair}}^{-1}\sum_{j=1}^{N_{\mathrm{pair}}}
 \nrmse(a^{\mathrm{est}}_j,a^{\mathrm{ref}}_j)}
 {N_{\mathrm{pair}}^{-1}\sum_{j=1}^{N_{\mathrm{pair}}}
 \nrmse(a^{\Raw{10}}_j,a^{\mathrm{ref}}_j)},
 \label{eq:aggregate-ratio-r14}
\end{equation}
where $N_{\mathrm{pair}}$ is the number of independent observation/reference pairs.  This quantity differs from the geometric mean of the pairwise ratios $r_{a,j}$ in \cref{eq:ratio-r14}.

\section{Geometry-native reconstruction and support monitoring}
\label{sec:method}

\subsection{Prior-plus-observation estimator}

Let $g$ identify a geometry and let $\mathcal T_g$ be a linear transform from native fields to geometry-adapted coefficients.  For a three-block observation, define $\boldsymbol z_3=\mathcal T_g\boldsymbol m^{\mathrm{obs}}_3$.  A geometry-specific coefficient prior $\overline{\boldsymbol z}_g$ and residual-transfer operator $\boldsymbol H_g$ give
\begin{equation}
 \widehat{\boldsymbol z}=\overline{\boldsymbol z}_g+
 \boldsymbol H_g(\boldsymbol z_3-\overline{\boldsymbol z}_g),
 \qquad
 \widehat{\boldsymbol m}^{\,\mathrm{pre}}=\mathcal T_g^{-1}\widehat{\boldsymbol z}.
 \label{eq:unified-r14}
\end{equation}
The superscript $\mathrm{pre}$ denotes the estimate before data-consistency projection.  The operator may be diagonal by spectral mode or a small matrix coupling related components.  Its singular values are restricted to $[0,1]$, so the observation residual is never amplified.  $\boldsymbol H_g=\boldsymbol 0$ returns the prior, whereas $\boldsymbol H_g=\boldsymbol I$ returns the observation; $\boldsymbol I$ is the identity operator.  The transform, prior, transfer coefficients, and masks are fixed before an evaluation observation is processed.

The transformed estimate is finally projected onto a declared data-consistency constraint.  For native fluid cells, define the area-weighted mean of a vector field $\boldsymbol a$ as
\begin{equation}
 \langle\boldsymbol a\rangle_A=
 \frac{\sum_c A_c\boldsymbol a_c}{\sum_c A_c}.
\end{equation}
Let $\boldsymbol q^{\mathrm{obs}}_3=(q^{\mathrm{obs}}_{x,3},q^{\mathrm{obs}}_{y,3})^{\mathsf T}$ and $\widehat{\boldsymbol q}^{\,\mathrm{pre}}=(\widehat q_x^{\,\mathrm{pre}},\widehat q_y^{\,\mathrm{pre}})^{\mathsf T}$ be the heat-flux subvectors of $\boldsymbol m^{\mathrm{obs}}_3$ and $\widehat{\boldsymbol m}^{\,\mathrm{pre}}$, respectively.  The cylinder estimate uses
\begin{equation}
 \widehat{\boldsymbol q}=\widehat{\boldsymbol q}^{\,\mathrm{pre}}+
 \left(\langle\boldsymbol q^{\mathrm{obs}}_3\rangle_A-
 \langle\widehat{\boldsymbol q}^{\,\mathrm{pre}}\rangle_A\right),
 \label{eq:mean-restore-r14}
\end{equation}
which exactly preserves the measured Cartesian mean.  The corresponding cavity constraint assigns unit weight to the zero-frequency coefficient.  These conditions do not assert that the three-block mean is exact; they prevent a historical prior from silently replacing a directly observed additive mode.

\subsection{Cavity estimator}

The cavity fields lie on a $100\times100$ Cartesian grid.  The structured prior is supplied by MambaIR, a deep image-restoration network built on the selective state-space (Mamba) sequence model \citep{GuDao2023Mamba,GuoEtAl2024MambaIR}.  A state-space layer carries a hidden state $\boldsymbol h_k$ along a pixel sequence $x_k$ through the discretised linear recurrence
\begin{equation}
 \boldsymbol h_k=\overline{\boldsymbol A}\,\boldsymbol h_{k-1}+\overline{\boldsymbol B}\,x_k,
 \qquad
 y_k=\boldsymbol C\,\boldsymbol h_k,
 \label{eq:ssm-r14}
\end{equation}
where $\overline{\boldsymbol A}=\exp(\Delta\boldsymbol A)$ and $\overline{\boldsymbol B}=(\Delta\boldsymbol A)^{-1}(\exp(\Delta\boldsymbol A)-\boldsymbol I)\,\Delta\boldsymbol B$ are zero-order-hold discretisations of the continuous system matrices $\boldsymbol A$ and $\boldsymbol B$ with step size $\Delta$, and $y_k$ is the layer output.  In the selective (Mamba) form, $\Delta$, $\boldsymbol B$, and $\boldsymbol C$ are functions of the current input rather than constants, so the recurrence can retain or discard information according to content while its cost remains linear in sequence length.  This combination of a global receptive field with linear cost is what makes a state-space backbone attractive for full-field restoration.  MambaIR adapts the layer to two-dimensional fields: the image is unfolded into four one-dimensional sequences (row-wise and column-wise, each in both directions), the selective scan is applied to each, and the four outputs are merged; residual state-space blocks then add a local convolution branch and channel attention, restoring the neighbouring-pixel interactions and reducing the channel redundancy that a purely sequential scan loses \citep{GuoEtAl2024MambaIR}.

No noise-free reference field exists for training, so the network is trained on the flow itself with the noise-to-noise principle: independent noisy realisations at the same operating condition serve as input/target pairs, and the minimiser of the resulting loss approaches the underlying mean field \citep{LehtinenEtAl2018Noise2Noise}.  An ensemble of such networks is trained, and its ensemble mean is the prior $q_y^{\mathrm{pri}}$ for the transverse heat flux.  The prior suppresses high-frequency particle noise but can shrink uncertain amplitudes or offsets toward its development distribution, so it is not used without a current-observation correction.

Let $Z^{\mathrm{pri}}_{kl}$ and $Z^{(3)}_{kl}$ be the two-dimensional discrete cosine transform (DCT) coefficients of the prior and three-block observation, respectively \citep{AhmedNatarajanRao1974DCT}; $k$ and $l$ are mode indices.  Development trajectories supply repeatable-signal power $\Phi^{\mathrm{sig}}_{kl}$ and single-block sampling-noise power $\Phi^{\mathrm{noise}}_{kl}$.  The final cavity estimate is
\begin{align}
 G_{kl} &= \frac{\Phi^{\mathrm{sig}}_{kl}}
 {\Phi^{\mathrm{sig}}_{kl}+\Phi^{\mathrm{noise}}_{kl}/3},
 \qquad G_{00}=1, \\
 \widehat Z_{kl} &= Z^{\mathrm{pri}}_{kl}+G_{kl}(Z^{(3)}_{kl}-Z^{\mathrm{pri}}_{kl}).
\label{eq:cavity-wiener-r14}
\end{align}
We refer to this estimator as \emph{Mamba--Wiener}: the prior coefficients $Z^{\mathrm{pri}}_{kl}$ come from the frozen MambaIR ensemble mean, and $G_{kl}\in[0,1]$ is the classical Wiener gain, the weight that minimises the expected squared error of a linear blend when signal and noise powers are known \citep{Wiener1949Filtering,ChungEtAl2011OptimalSpectralFilters}.  Each mode therefore leans on the observation exactly to the degree that development data say it is trustworthy: $G_{kl}\to 1$ where repeatable signal dominates, $G_{kl}\to 0$ where three blocks are mostly noise.  The division by three converts single-block noise power to that of the three-block average, and $G_{00}=1$ keeps the measured spatial mean untouched.

For the comparisons reported later (\cref{tab:cavity-r14}), the alternative estimators are defined once here.  They serve three purposes: anchoring the sampling budgets, testing whether a generic single-realisation filter could deliver the same gain, and attributing the gain of \cref{eq:cavity-wiener-r14} between its prior and its observation.  $\mathrm{Raw}(B)$ is the direct $B$-block average with no processing; \Raw{3} and \Raw{10} fix the low-budget and comparator anchors.  TSVD/POD represents the generic filter class: it projects $\mathrm{Raw}(3)$ onto its leading $r$ singular modes, $\widehat q=\sum_{i\le r}\sigma_i u_i v_i^{\mathsf T}$, with $r$ fixed on development data, and discards the noise-dominated tail \citep{BerkoozEtAl1993POD}.  It uses no historical information, so any margin of the final estimator over it must come from the learned prior or from the observation-conditioned gain.  The remaining rows are ablations of \cref{eq:cavity-wiener-r14} itself.  The \emph{Mamba prior} row sets $G_{kl}\equiv 0$, giving $\widehat Z_{kl}=Z^{\mathrm{pri}}_{kl}$: the frozen network output with the current observation ignored, which measures how far historical structure alone carries.  \emph{Zero-frequency restoration only} keeps $G_{kl}\equiv 0$ except $G_{00}=1$, so the prior is merely re-centred on the measured spatial mean; its gap to the full estimator isolates the value of the nonzero observed modes.  The \emph{bounded data-consistent estimate} is the intermediate stage of \cref{eq:unified-r14} under the cavity zero-frequency constraint, the prior corrected by the bounded observation residual before the continuous per-mode Wiener allocation, so its gap to the final row measures that allocation.  The \emph{cross-condition permutation} feeds \cref{eq:cavity-wiener-r14} an observation from the wrong flow condition; if generic smoothing or the gain structure alone explained the improvement, this row would remain competitive, and its failure rules that explanation out.  Because the final gain allocation was selected within the cavity analysis, all of these results are treated as method development rather than independent confirmation.

\subsection{Cylinder-native estimator}

Let $(x,y)$ be a native fluid-cell centre, $(x_c,y_c)$ the cylinder centre, $R_c$ the radius, $D=2R_c$ the diameter, and
\begin{equation}
 r_c=\sqrt{(x-x_c)^2+(y-y_c)^2},
 \qquad
 \cos\theta=\frac{x-x_c}{r_c},\quad
 \sin\theta=\frac{y-y_c}{r_c}.
\end{equation}
The angle $\theta$ is measured counter-clockwise from the downstream $+x$ direction.  Define the outward unit normal $\boldsymbol n=(\cos\theta,\sin\theta)^{\mathsf T}$ and counter-clockwise unit tangent $\boldsymbol t=(-\sin\theta,\cos\theta)^{\mathsf T}$.  Cartesian heat flux is rotated into $q_n=\boldsymbol q\cdot\boldsymbol n$ and $q_t=\boldsymbol q\cdot\boldsymbol t$,
\begin{equation}
 q_n=q_x\cos\theta+q_y\sin\theta,
 \qquad
 q_t=-q_x\sin\theta+q_y\cos\theta.
 \label{eq:rotate-r14}
\end{equation}
Only fluid cells enter the transform.  The components are interpolated to a $128\times96$ grid in $(r_c/D,\theta)$ and represented by a two-dimensional DCT.  Native areas are retained for \cref{eq:nrmse-r14,eq:mean-restore-r14}.

Four development trajectories define the cylinder prior $\overline{\boldsymbol z}_g$ as the mean of their ten-block coefficient fields.  For distinct trajectories $i$ and $j$, $\boldsymbol z_i^{(3)}$ and $\boldsymbol z_i^{(10)}$ denote their three- and ten-block coefficient vectors.  Define the low-budget and target peer residuals $\boldsymbol d^{\mathrm{low}}_{ij}$ and $\boldsymbol d^{\mathrm{tar}}_{ij}$ by
\begin{equation}
 \boldsymbol d^{\mathrm{low}}_{ij}=\boldsymbol z_i^{(3)}-\boldsymbol z_j^{(10)},
 \qquad
 \boldsymbol d^{\mathrm{tar}}_{ij}=\boldsymbol z_i^{(10)}-\boldsymbol z_j^{(10)}.
\end{equation}
The radial DCT modes $k=0,\ldots,127$ are divided into four contiguous groups with edges $(0,32,64,96,128)$, and the angular modes $l=0,\ldots,95$ into groups with edges $(0,24,48,72,96)$.  Within each resulting bin $b$ of this fixed $4\times4$ partition, a two-component transfer matrix is obtained from
\begin{equation}
 \widetilde{\boldsymbol H}_b=
 \operatorname*{arg\,min}_{\boldsymbol H}
 \sum_{i\ne j}\|\boldsymbol d^{\mathrm{tar}}_{ij,b}-\boldsymbol H\boldsymbol d^{\mathrm{low}}_{ij,b}\|_2^2
 +\lambda_b\|\boldsymbol H\|_F^2,
 \label{eq:ridge-r14}
\end{equation}
where $\|\cdot\|_2$ and $\|\cdot\|_F$ are the Euclidean and Frobenius norms.  If $\boldsymbol D_b$ is the matrix whose columns are the two-component samples in $\boldsymbol d^{\mathrm{low}}_{ij,b}$, the fixed half-trace scaling is $\lambda_b=\tfrac12\operatorname{tr}(\boldsymbol D_b\boldsymbol D_b^{\mathsf T})$.  If $\widetilde{\boldsymbol H}_b=\boldsymbol U\boldsymbol\Sigma\boldsymbol V^{\mathsf T}$ is its singular-value decomposition, $\boldsymbol U$ and $\boldsymbol V$ contain the left and right singular vectors and $\boldsymbol\Sigma$ contains the singular values.  Each diagonal entry of $\boldsymbol\Sigma$ is clipped to $[0,1]$ before reconstructing $\boldsymbol H_b$.  The fitted nonzero-mode singular values range from 0.0882 to 0.1630, implying strong suppression of observation-dominated fluctuations.  The zero-frequency matrix is the identity, followed by the native Cartesian correction in \cref{eq:mean-restore-r14}.

For a new cylinder observation, \cref{eq:unified-r14} is applied binwise to $(q_n,q_t)$, followed by inverse transformation, rotation to $(q_x,q_y)$, and native-cell data consistency.  A prior-only result identifies the contribution of repeatable historical structure.  A phase-scrambled residual retains the same spectral amplitudes but destroys alignment with the shock, near-wall layer, and wake; it tests whether the present observation contributes meaningful spatial information.

\subsection{Complete-field support score}

Operating-condition support is evaluated from the current observation, not from its error against a reference.  A linearly extrapolated Mach-12 prior is formed from development mean fields,
\begin{equation}
 \overline{\boldsymbol m}_{12}^{\,\mathrm{ext}}
 =2\overline{\boldsymbol m}_{10}-\overline{\boldsymbol m}_{8},
 \label{eq:mach-prior-r14}
\end{equation}
where $\overline{\boldsymbol m}_8$ and $\overline{\boldsymbol m}_{10}$ denote the Mach-8 and Mach-10 mean fields.  The monitored components are $(n,u,v,T,P_{xx},P_{xy},P_{yy},q_n,q_t)$.  For component index $\ell$, monitor-zone index $\zeta$, and evaluation trajectory $j$, an empirical-Bayes observation gain is
\begin{equation}
 K^{(j)}_{\ell\zeta}=
 \operatorname{clip}_{[0,1]}\!\left(
 1-\frac{\sigma^2_{\ell\zeta}}
 {V^{\mathrm{res},(j)}_{\ell\zeta}}
 \right),
 \label{eq:eb-r14}
\end{equation}
where $\operatorname{clip}_{[0,1]}(x)=\min(1,\max(0,x))$.  With $w_c=A_c/\sum_{c\in\zeta}A_c$, the current residual power is
\begin{equation}
 V^{\mathrm{res},(j)}_{\ell\zeta}
 =\sum_{c\in\zeta}w_c
 \left(m^{\mathrm{obs},(j)}_{3,\ell c}
 -\overline m^{\,\mathrm{ext}}_{12,\ell c}\right)^2.
\end{equation}
The quantity $\sigma^2_{\ell\zeta}$ is the corresponding area-weighted within-trajectory sampling-noise power estimated from the four three-block development draws.  The monitor masks are: near wall, $r_c\le0.20\,\mathrm m$; wake, $r_c>0.20\,\mathrm m$ and $x-x_c\ge0$; and outer flow, $r_c>0.20\,\mathrm m$ and $x-x_c<0$.  These masks define the three values of $\zeta$.

Development data define the largest supported gain $K^{\max,\mathrm{dev}}_{\ell\zeta}$ for each of the $9\times3=27$ component--zone combinations.  With $\mathbf 1[\cdot]$ denoting the indicator function, the deployment outside-envelope fraction is
\begin{equation}
 s_j=\frac{1}{27}\sum_{\ell,\zeta}
 \mathbf 1\!\left[K^{(j)}_{\ell\zeta}>
 K^{\max,\mathrm{dev}}_{\ell\zeta}\right].
 \label{eq:support-r14}
\end{equation}
For reporting the separation from the envelope, the maximum gain-to-envelope ratio is
\begin{equation}
 R_j^{K}=\max_{\ell,\zeta}
 \frac{K^{(j)}_{\ell\zeta}}{K^{\max,\mathrm{dev}}_{\ell\zeta}},
 \label{eq:gain-envelope-ratio-r14}
\end{equation}
whose denominator is positive for every monitored component--zone combination in the development data.
Calibration preserves the four three-block draws available within each development trajectory.  For development unit $i$, draw $d\in\{1,\ldots,4\}$, let $K^{(i,d)}_{\ell\zeta}$ be the gain computed from that draw and let $K^{\max,(-i)}_{\ell\zeta}$ be the envelope constructed without unit $i$.  Define
\begin{equation}
 \widetilde s_i=\frac{1}{4\times27}
 \sum_{d=1}^{4}\sum_{\ell,\zeta}
 \mathbf 1\!\left[K^{(i,d)}_{\ell\zeta}>
 K^{\max,(-i)}_{\ell\zeta}\right].
 \label{eq:support-calibration-r14}
\end{equation}
The maximum of the $N_{\mathrm{dev}}=8$ leave-one-unit-out values in \cref{eq:support-calibration-r14} is used as the conservative development-envelope threshold, $s_\star=47/108=0.435185$.  The same numerical threshold is applied to the single-draw deployment score in \cref{eq:support-r14}: observation $j$ is accepted when $s_j\le s_\star$ and otherwise receives an abstention.  Because the calibration and deployment scores aggregate different numbers of draws, this maximum-score rule is treated as a fixed development heuristic, not as a finite-sample miscoverage guarantee.  The construction borrows the logic of split-conformal calibration \citep{Vovk2005ALRW,ShaferVovk2008Tutorial,AngelopoulosBates2023CP}: the threshold comes from development scores alone, although the score aggregated here is familywise over component--zone gains rather than a pointwise nonconformity measure.

A separate same-condition Mach-12 analysis measures reconstructability when in-condition information is available.  Two 40-block reference trajectories define the pointwise prior $\overline m_{\ell c}$ for component $\ell$ and native fluid cell $c$.  A distinct attribution map $\eta(c)$ has three zones: near cylinder, $r_c\le0.20\,\mathrm m$; outer downstream, $r_c>0.20\,\mathrm m$ and $x-x_c\ge0$; and outer upstream, $r_c>0.20\,\mathrm m$ and $x-x_c<0$.  For the remaining trajectories,
\begin{equation}
 \widehat m_{\ell c}=\overline m_{\ell c}+
 \beta_{\ell,\eta(c)}
 \left(m^{(3)}_{\ell c}-\overline m_{\ell c}\right),
 \qquad 0\le\beta_{\ell,\eta(c)}\le1,
 \label{eq:same-condition-r14}
\end{equation}
where $m^{(3)}_{\ell c}$ is the three-block observation and $\beta_{\ell,\eta}$ is a component--zone shrinkage coefficient.  The coefficients are fitted by alternating the two calibration trajectories as prior and target, using 20 fixed three-block draws and the complementary 37 blocks as target, and clipping the resulting least-squares slopes to $[0,1]$.  Heat flux is fitted pointwise in $(q_n,q_t)$ and rotated back to Cartesian components.  This attribution analysis is kept separate from the extrapolated support decision.

\section{Numerical setup and validation design}
\label{sec:design}

\subsection{Flow configurations and sampling}

\Cref{fig:workflow-r14} summarises the estimator and its comparison data.  The three-block observation is the only current-trajectory input to reconstruction.  A ten-block direct average is a paired comparator, and an independently generated trajectory or leave-one-seed-out average provides the reference.  The cavity uses a nested three/ten-block comparison, whereas the cylinder uses disjoint three- and ten-block windows.  Additive accumulators are combined according to \cref{eq:additive-r14,eq:central-third-r14} on every branch.

\begin{journalfigurepage}
  \journalgraphic{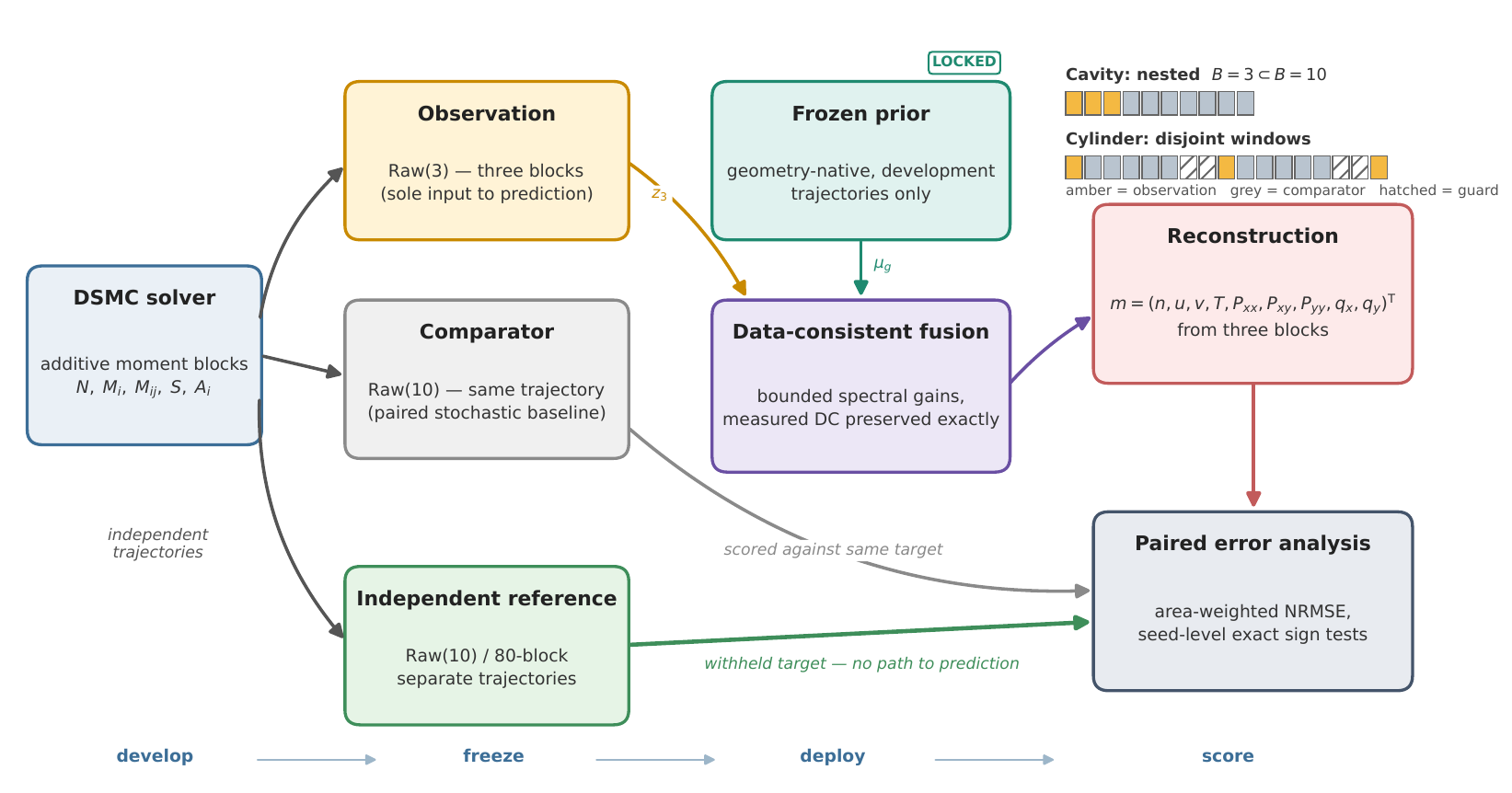}
  \caption{Prediction and evaluation dataflow.  Additive DSMC blocks are selected as a nested three/ten-block comparison for the cavity and as disjoint three- and ten-block windows for the cylinder.  The observation and a fixed geometry-native prior produce the reconstructed field vector $\boldsymbol m=(n,u,v,T,P_{xx},P_{xy},P_{yy},q_x,q_y)^{\mathsf T}$.  Reconstruction and the comparator are scored only against the independent reference, which has no path to prediction.}
  \label{fig:workflow-r14}
\end{journalfigurepage}

\begin{journalfigurepage}
  \journalgraphic{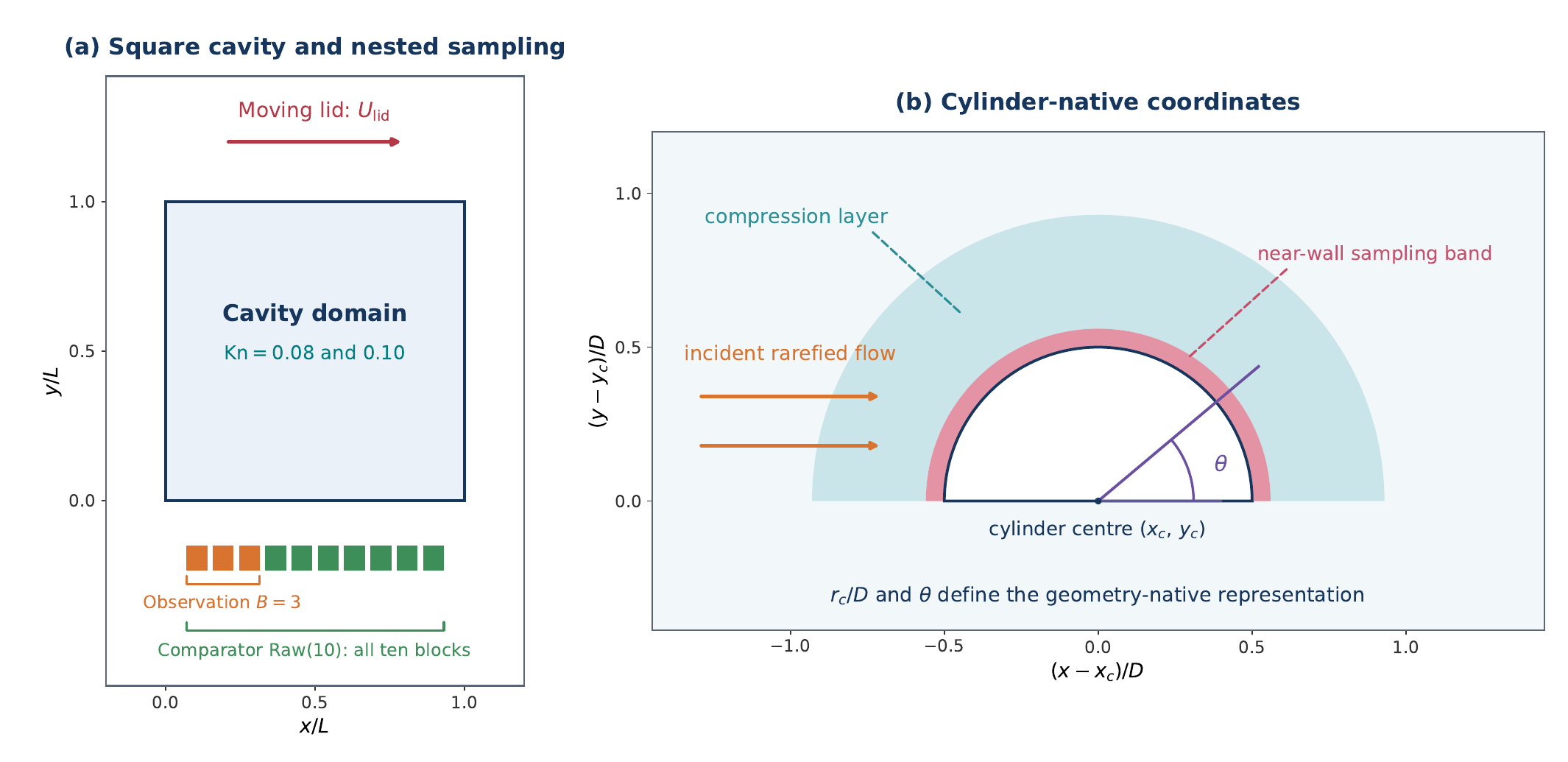}
  \caption{Computational geometries and sampling coordinates.  (a) Square cavity of side $L$ with lid speed $U_{\mathrm{lid}}$, Knudsen number $\mathrm{Kn}$, and dimensionless coordinates $(x/L,y/L)$.  The block icons distinguish observation and comparator roles; the cavity three-block set is nested within its ten-block comparator.  (b) Circular cylinder of diameter $D$, centred at $(x_c,y_c)$, with dimensionless radial coordinate $r_c/D$ and polar angle $\theta$ measured counter-clockwise from downstream.  The shaded annuli denote the nearest-cell evaluation band and the compression layer.}
  \label{fig:geometries-r14}
\end{journalfigurepage}

The Knudsen number is $\mathrm{Kn}=\lambda_{\mathrm{ref}}/L_c$, where $\lambda_{\mathrm{ref}}$ is the reference mean free path, the equilibrium reference-state value for the cavity and the freestream value for the cylinder, and the characteristic length $L_c$ is the cavity side $L$ or cylinder diameter $D$.  The cavity is a $100\times100$ Cartesian DSMC calculation of monatomic argon.  Four independent seeds are analysed at each of two conditions: $(\mathrm{Kn},U_{\mathrm{lid}})=(0.08,350\,\mathrm{m\,s^{-1}})$ and $(0.10,400\,\mathrm{m\,s^{-1}})$.  Each trajectory contains ten additive sampling blocks.  Blocks 0--2 define the low-budget observation, and all ten define its direct comparator.  The reference for a seed is the leave-one-seed-out mean of the ten-block fields from the other three seeds at the same condition; the evaluated seed therefore contributes no samples to its reference.  Two trajectories did not meet a conservative temperature-extremum stationarity diagnostic, although their mechanical, finiteness, collision, and heat-flux-specific checks were retained.  Together with in-condition gain selection, this caveat is why the cavity study is used for method development rather than confirmatory inference.

The cylinder calculations use Bird's DS2V visual DSMC program for two-dimensional and axisymmetric flows, operated here in its two-dimensional mode, on native adaptive cells.  The gas is argon at $\mathrm{Kn}=0.1$, freestream temperature $T_\infty=200\,\mathrm K$, and diffuse wall temperature $T_w=500\,\mathrm K$.  The freestream Mach number is $M=10$ and the corresponding speed is $U_\infty=2634.1\,\mathrm{m\,s^{-1}}$.  Four development trajectories define the geometry-native model.  The independent evaluation comprises six observation/reference pairs, with all 12 seeds distinct from development.  For each observation, solver-output indices 100, 108, and 116 define $\Raw{3}$; ten different indices, 101--105 and 109--113, define the paired $\Raw{10}$ comparator.  The same ten indices on the partner trajectory define the independent reference.  Four intervening outputs are unused guards.  The evaluated window spans $11.268\le tU_\infty/D\le11.390$, where $tU_\infty/D$ is convective time; conclusions are restricted to this finite-time window.

The support monitor uses four Mach-8 and four Mach-10 development trajectories, each containing 40 blocks.  Held-out specificity is measured on the six Mach-10 observations just described, without their reference partners.  Four additional Mach-12 observations are then processed by the fixed rule.  Each contains at least 40 complete outputs; the final 14 define a three-block observation, one guard, and a disjoint ten-block comparator.  Independent Mach-12 references are withheld until after classification.  The monitor's form was motivated in part by earlier Mach-12 experience, so the out-of-support evidence is restricted to its performance on these four independent observations and is not presented as the discovery of a universal rule.

A separate same-condition Mach-12 analysis uses 12 evaluation trajectories and four 40-block reference trajectories.  Two references define the prior and shrinkage coefficients in \cref{eq:same-condition-r14}; the other two form an 80-block verification target.  Temporal-correlation statistics pool all four reference trajectories, whereas accuracy is evaluated only against the 80-block target.  This split quantifies reconstruction when same-condition support is explicitly supplied and is not used to tune the extrapolated support decisions.

The numerical specifications required to interpret the reported comparisons are collected in \cref{tab:setup-r14}.  The variable-hard-sphere (VHS) and variable-soft-sphere (VSS) models describe molecular scattering; no-time-counter (NTC) denotes the collision-pair selection algorithm.

\begin{table}[p]
\centering
\small
\caption{Physical and numerical specifications.  $M$ is freestream Mach number and $n_\infty$ is freestream number density.  The DS2V solver uses local adaptive time steps, so no single global time increment is reported.}
\label{tab:setup-r14}
\begin{tabularx}{\textwidth}{@{}>{\raggedright\arraybackslash}p{0.22\textwidth}Y Y@{}}
\toprule
Item & Cavity & Circular cylinder \\
\midrule
Solver and gas model & Cartesian DSMC; monatomic argon; VHS scattering; NTC collisions & Bird's DS2V; monatomic argon; nearest-neighbour collisions; viscosity exponent 0.74; reciprocal VSS parameter 1.0 \\
Geometry and cells & Square of side $L$; $100\times100$ equal-area cells & $D=0.3048\,\mathrm m$; domain $x\in[-0.20,0.65]\,\mathrm m$, $y\in[0,0.40]\,\mathrm m$; 17,312 populated native cells \\
Operating conditions & $\mathrm{Kn}=0.08$ and 0.10; $U_{\mathrm{lid}}=350$ and $400\,\mathrm{m\,s^{-1}}$ & $\mathrm{Kn}=0.1$; $M=8,10,12$; $U_\infty=2107.28,2634.10,3160.92\,\mathrm{m\,s^{-1}}$; $T_\infty=200\,\mathrm K$; $n_\infty=4.247\times10^{20}\,\mathrm{m^{-3}}$ at $M=12$ \\
Molecular parameters & $m_{\mathrm{mol}}=6.63\times10^{-26}\,\mathrm{kg}$; reference diameter $4.17\times10^{-10}\,\mathrm m$ at $273\,\mathrm K$; VHS exponent 0.81 & $m_{\mathrm{mol}}=6.63\times10^{-26}\,\mathrm{kg}$; reference diameter $3.595\times10^{-10}\,\mathrm m$ at $1000\,\mathrm K$ \\
Wall and particles & Fully diffuse wall; $T_w=300\,\mathrm K$; initially 32 simulator particles per cell & Fully diffuse wall; $T_w=500\,\mathrm K$; $1.5\times10^6$ initial simulator particles \\
Sampling & 300 samples per block; ten retained blocks in each analysed trajectory & Additive volume moments and a native wall-collision energy tally at each retained output; partitions defined above \\
\bottomrule
\end{tabularx}
\end{table}

\subsection{Endpoints and inference}

The cavity endpoint is the equal-area NRMSE of transverse heat flux $q_y$.  The Mach-10 co-primary endpoints are the area-weighted NRMSE of $q_y$ over all fluid cells and of normal heat flux $q_n$ over cells satisfying $0\le r_c-R_c\le0.05D$.  Both the selected estimate and paired ten-block comparator are evaluated against the same independent reference.  The result is reported as the ratio of arithmetic-mean NRMSEs in \cref{eq:aggregate-ratio-r14} and as the six individual ratios in \cref{eq:ratio-r14}.

For each endpoint, six improvements among $N_{\mathrm{pair}}=6$ independent pairs have one-sided exact sign-test probability $2^{-6}=0.015625$.  Holm's step-down correction across the two co-primary endpoints gives 0.03125 when both raw probabilities are equal \citep{Holm1979}.  A geometric mean of pair ratios and a descriptive Student-$t$ interval on their logarithms are also reported.  The exact sign test is the inferential result; native cells are not treated as replicates.  With only $N_{\mathrm{seed}}=4$ cavity seeds per condition, the minimum one-sided exact probability is 0.0625, reinforcing its developmental interpretation.
For the same-condition Mach-12 analysis, each displayed point is the geometric mean of 12 trajectory-level ratios.  Its 95\% percentile interval is formed from 20,000 bootstrap resamples of the 12 log ratios, with trajectories resampled as complete units and the geometric mean recomputed for each resample.

Support performance is evaluated by acceptance of at least five of the six held-out Mach-10 observations and abstention on all four independent Mach-12 observations.  Reference-derived error is not part of either classification.  Independent-reference scoring is reported separately as a diagnostic of how conservative the support rule is.

Temporal dependence is quantified by the integrated autocorrelation time
\begin{equation}
 \tau_{\mathrm{int}}=1+2\sum_{k=1}^{k^\star}\rho_k,
 \qquad B_{\mathrm{eff}}=\frac{B}{\tau_{\mathrm{int}}},
 \label{eq:beff-r14}
\end{equation}
where $\rho_k$ is the pooled lag-$k$ block autocorrelation, $k^\star$ is the last lag before the first non-positive value, and $B_{\mathrm{eff}}$ is the effective number of independent blocks among $B$ nominal blocks.  The cell-centred quantity $q_n$ is distinguished from the native wall-collision heat flux $q_w$, defined as net molecular energy delivered to the wall per unit area and time.

For the Mach-12 wall statistic, sampling-scale oscillation near the upstream stagnation point $\theta=\pi$ is represented by the symmetry-constrained local model
\begin{equation}
 \widetilde q(\theta)=a+b(\pi-\theta)^2,
 \label{eq:stagnation-r14}
\end{equation}
where $\theta$ is in radians.  The coefficients $a$ and $b$ are fitted separately to $q_w$ and $-q_n$ over $149^{\circ}\le\theta\le180^{\circ}$ by inverse-variance weighted least squares.  Between $149^{\circ}$ and $159^{\circ}$, the raw mean and fitted trend are joined by the smoothstep weight $h(\xi)=\xi^2(3-2\xi)$, where $\xi=(\theta-149^{\circ})/10^{\circ}$; the fit has full weight above $159^{\circ}$.  It enforces $\partial\widetilde q/\partial\theta=0$ at stagnation, and the raw bin values remain reported alongside the fit.  The displayed 95\% sampling bands are $\overline q(\theta)\pm1.96s_q(\theta)/\sqrt{B_{\mathrm{eff}}}$, where $\overline q(\theta)$ and $s_q(\theta)$ are the angular-bin block mean and standard deviation.  They use the held-out 80-block pool, for which $B_{\mathrm{eff}}=6.5198$ for $q_w$ and 54.6327 for the angular $-q_n$ profile.

\section{Results}
\label{sec:results}

\subsection{Cavity reconstruction}
\label{sec:cavity-results-r14}

The cavity calculations isolate the central statistical difficulty.  The learned prior reproduces the organisation of $q_y$ but can retain a coherent amplitude and offset error, whereas \Raw{3} contains the current low-frequency content together with much larger sampling noise.  Enforcing the scalar identity in \cref{eq:energy-r14} reduces its residual but cannot determine the solenoidal part of the heat-flux error in \cref{eq:null-r14}.  The reconstruction must therefore preserve coherent structure while returning observation-supported modes to the field.

The equal-area errors in \cref{tab:cavity-r14} quantify this trade-off.  The final Mamba--Wiener estimator has mean NRMSE ratios of 0.6575 and 0.6723 relative to the paired \Raw{10} comparator at $(\mathrm{Kn},U_{\mathrm{lid}})=(0.08,350\,\mathrm{m\,s^{-1}})$ and $(0.10,400\,\mathrm{m\,s^{-1}})$, respectively.  Three observation blocks therefore reduce the error by roughly one third relative to ten direct blocks at both conditions.  The prior alone is unreliable at the second condition, where its ratio is 1.8861; the continuous Wiener residual corrects the broad bias without restoring the full high-frequency variance of \Raw{3}.  Cross-condition permutation gives ratios above 3.27, which rules out generic smoothing as the explanation.  Because the gain is estimated within the same two-condition analysis and only four seeds are available per condition, the cavity evidence is interpreted as method development rather than confirmatory inference.

The spatial diagnostics point in the same direction and show the mechanism behind the table.  In \cref{fig:cavity-ensemble-r14}, the reference and reconstruction share the circulation-induced structure of $q_y$, the residual bias is smooth rather than speckled, and positive root-mean-square-error (RMSE) reduction occupies most of the active region.  The representative fields in \cref{fig:cavity-008-r14,fig:cavity-010-r14} explain why the estimators in \cref{tab:cavity-r14} behave so differently.  \Raw{3} is dominated by fine-grained sampling texture: heat flux is a signed third-order moment whose large molecular contributions cancel, so its cell estimates carry the largest relative sampling noise in the retained hierarchy, and three blocks leave that variance essentially undiluted.  The Mamba prior removes this texture almost completely, which is the expected behaviour of a noise-to-noise network: trained to map one noisy realisation to another, its loss minimiser approaches the mean of the development fields, and the ensemble average suppresses the remaining seed-to-seed variability.  The same training objective is also the source of its failure at the second condition: a network fitted to the development distribution pulls the broad low-frequency amplitude of an individual condition toward that distribution, so the prior stays visually convincing while carrying the coherent amplitude and offset error quantified in the table.  The bias it leaves behind is smooth by construction, and smooth content is precisely what the current observation still measures with a usable signal-to-noise ratio at low mode numbers.  The independently sampled \Raw{10} field remains stochastic and is a finite-budget comparator, not exact truth.

The profiles in \cref{fig:cavity-vertical-r14,fig:cavity-horizontal-r14} localise the correction physically.  Through the near-lid layer, where the driven shear generates the strongest transverse transport, and across the interior recirculation, the conditioned estimate follows the reference amplitude that the prior underestimates, without reintroducing the block-to-block variance of \Raw{3}.  The same profiles indicate why the generic controls in \cref{tab:cavity-r14} cannot close this gap.  TSVD/POD retains the most energetic singular modes of one noisy realisation, but at three blocks those leading modes already mix sampling noise into the signal, and the truncation discards weak large-scale content that a cancellation-dominated signed moment needs.  Zero-frequency restoration corrects the single number its constraint protects, the spatial mean, and leaves the spatially varying low-mode amplitude error untouched.

The DCT gain in \cref{fig:cavity-spectrum-r14} makes the allocation explicit.  Development data assign $G_{kl}\to1$ only in a compact low-$(k,l)$ region where repeatable signal power exceeds three-block noise power, and $G_{kl}\to0$ elsewhere, so the estimator is a per-mode bias--variance compromise: prior-based variance suppression at high frequencies and observation-anchored amplitude at low frequencies.  The map also explains the severity of the cross-condition permutation: a permuted observation injects the wrong condition's content exactly in the trusted low-frequency band, which is why its error exceeds even \Raw{3} and why the gain structure alone cannot account for the improvement.

\begin{table}[htbp]
\centering
\small
\caption{Cavity transverse-heat-flux accuracy.  Entries are condition-mean $q_y$ NRMSE ratios to the paired \Raw{10} comparator; lower values indicate smaller error against the leave-one-seed-out reference.  Each condition contains four independent evaluated seeds.}
\label{tab:cavity-r14}
\begin{tabularx}{\textwidth}{@{}Yrr@{}}
\toprule
Estimator & \shortstack{$\mathrm{Kn}=0.08$\\$U_{\mathrm{lid}}=350\,\mathrm{m\,s^{-1}}$} & \shortstack{$\mathrm{Kn}=0.10$\\$U_{\mathrm{lid}}=400\,\mathrm{m\,s^{-1}}$} \\
\midrule
\Raw{3} & 1.6502 & 1.6560 \\
Mamba prior & 1.1010 & 1.8861 \\
TSVD/POD control & 0.9675 & 1.1223 \\
Zero-frequency restoration only & 0.8720 & 1.1594 \\
Bounded data-consistent estimate & 0.7506 & 0.9542 \\
Mamba--Wiener estimate & \textbf{0.6575} & \textbf{0.6723} \\
Cross-condition permutation & 3.2751 & 3.2932 \\
\Raw{10} & 1.0000 & 1.0000 \\
\bottomrule
\end{tabularx}
\end{table}

\begin{journalfigurepage}
  \journalgraphic{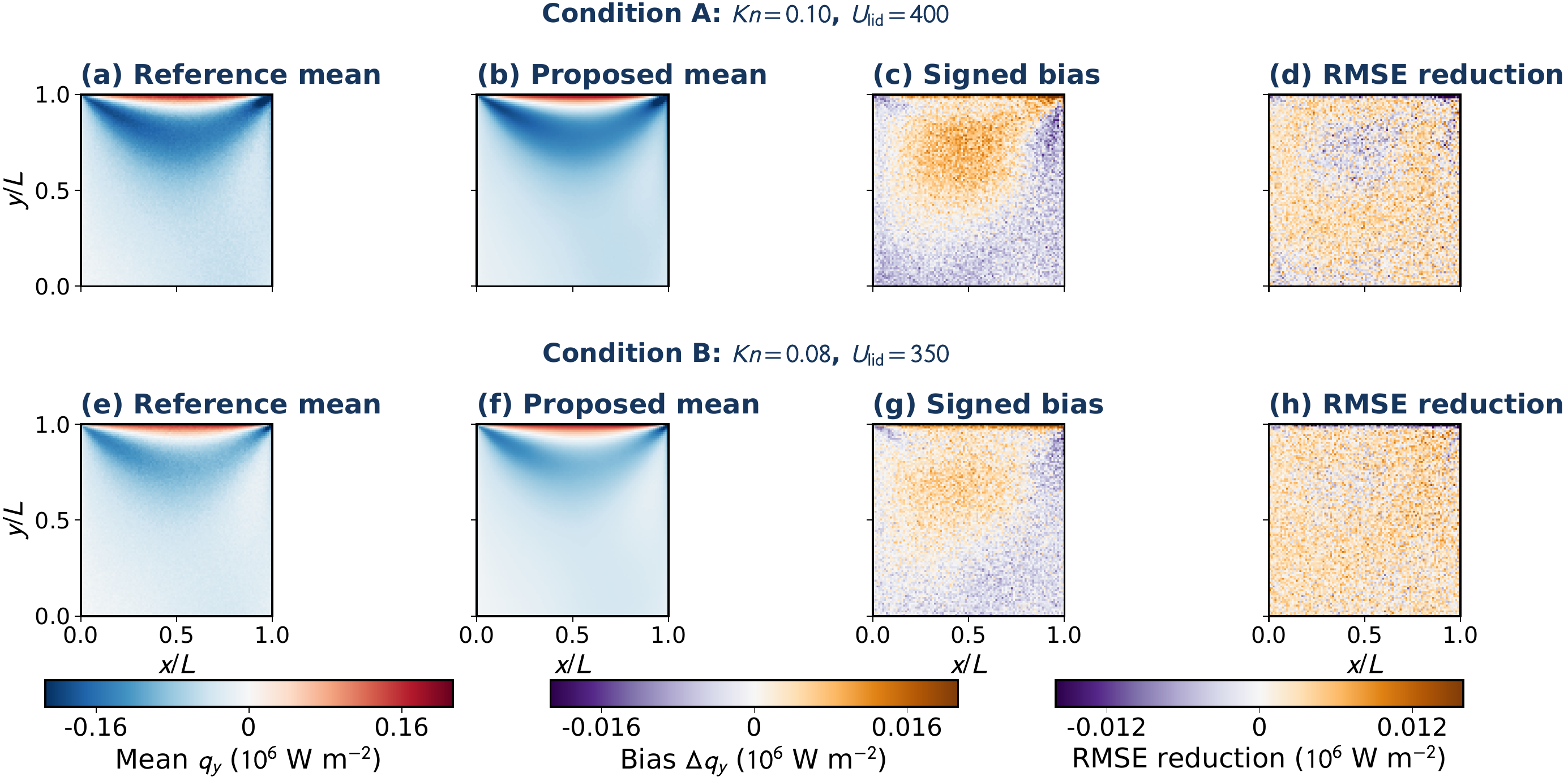}
  \caption{Cavity ensemble diagnostics at both conditions.  The plotted variable is transverse heat flux $q_y$ in $\mathrm{W\,m^{-2}}$.  Columns show the independent-reference mean, reconstructed mean, signed bias, and local $\Delta\mathrm{RMSE}=\mathrm{RMSE}_{B=10}-\mathrm{RMSE}_{\mathrm{rec}}$.  Positive $\Delta\mathrm{RMSE}$ denotes improvement over \Raw{10}; four independent seeds enter each row.}
  \label{fig:cavity-ensemble-r14}
\end{journalfigurepage}
\begin{journalfigurepage}
  \journalgraphic{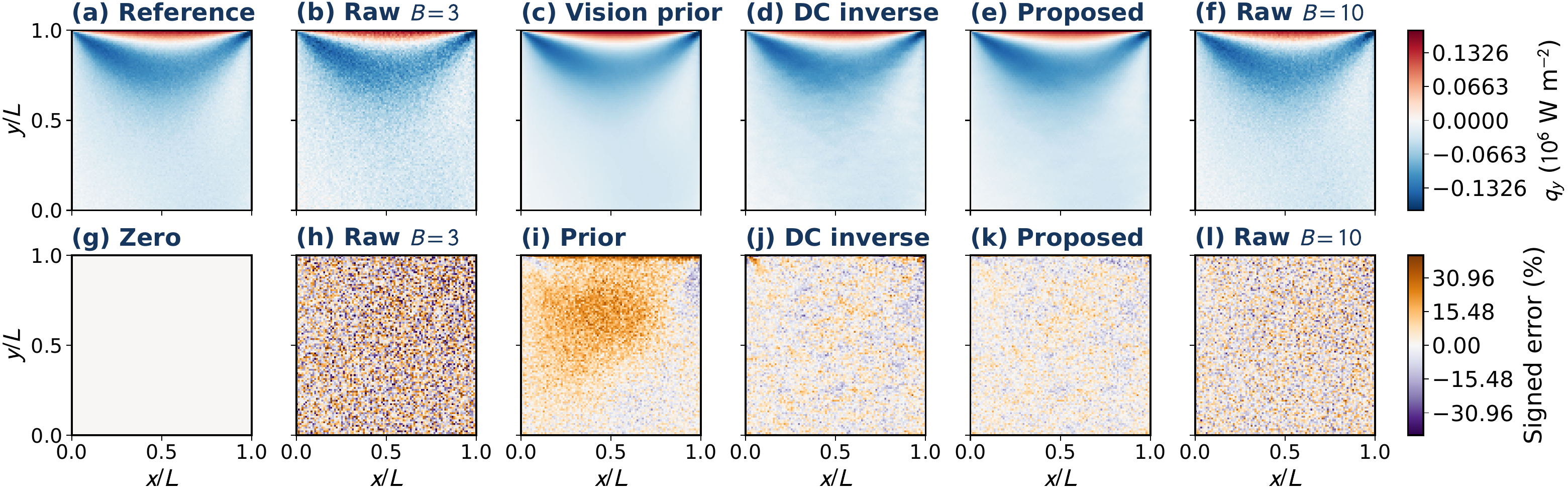}
  \caption{Cavity $q_y$ at $\mathrm{Kn}=0.08$ and $U_{\mathrm{lid}}=350\,\mathrm{m\,s^{-1}}$.  The upper row shows the leave-one-seed-out reference, \Raw{3}, Mamba prior, bounded data-consistent estimate, Mamba--Wiener estimate, and paired \Raw{10}.  The lower row gives signed error normalised by the reference root-mean-square magnitude; colour limits are identical within each row.}
  \label{fig:cavity-008-r14}
\end{journalfigurepage}
\begin{journalfigurepage}
  \journalgraphic{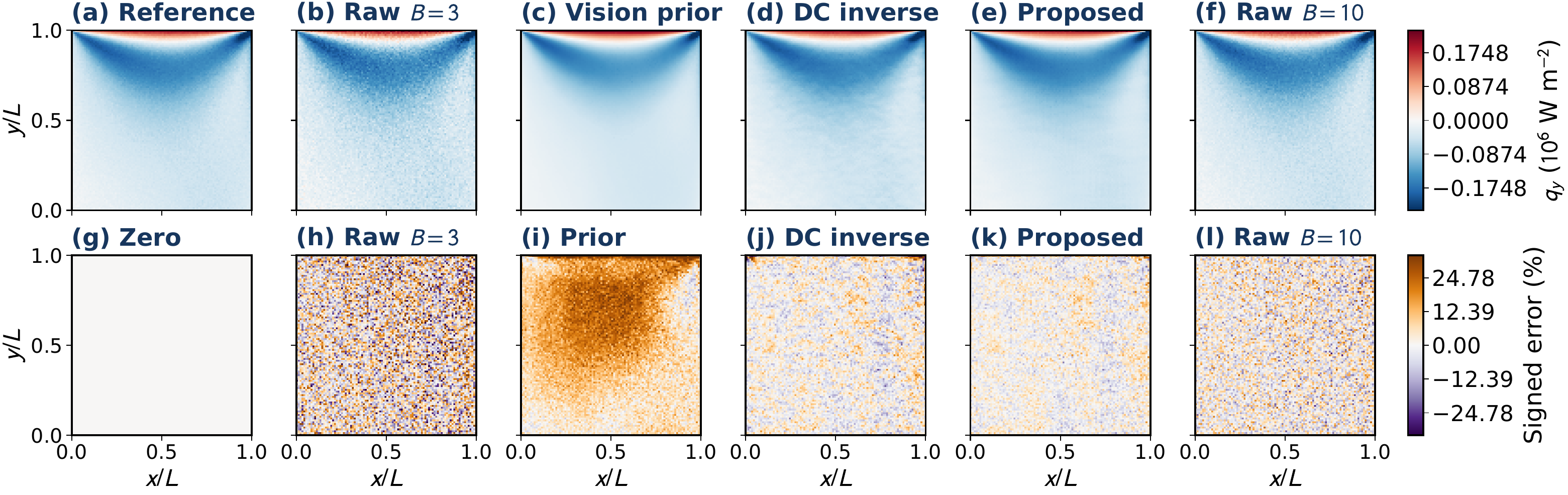}
  \caption{Cavity $q_y$ at $\mathrm{Kn}=0.10$ and $U_{\mathrm{lid}}=400\,\mathrm{m\,s^{-1}}$.  Panel order, normalisation, and within-row colour limits are those of \cref{fig:cavity-008-r14}.  The prior recovers the spatial pattern but underestimates broad low-frequency content; conditioning on the observation reduces this bias without reintroducing the variance of \Raw{3}.}
  \label{fig:cavity-010-r14}
\end{journalfigurepage}
\begin{journalfigurepage}
  \journalgraphic{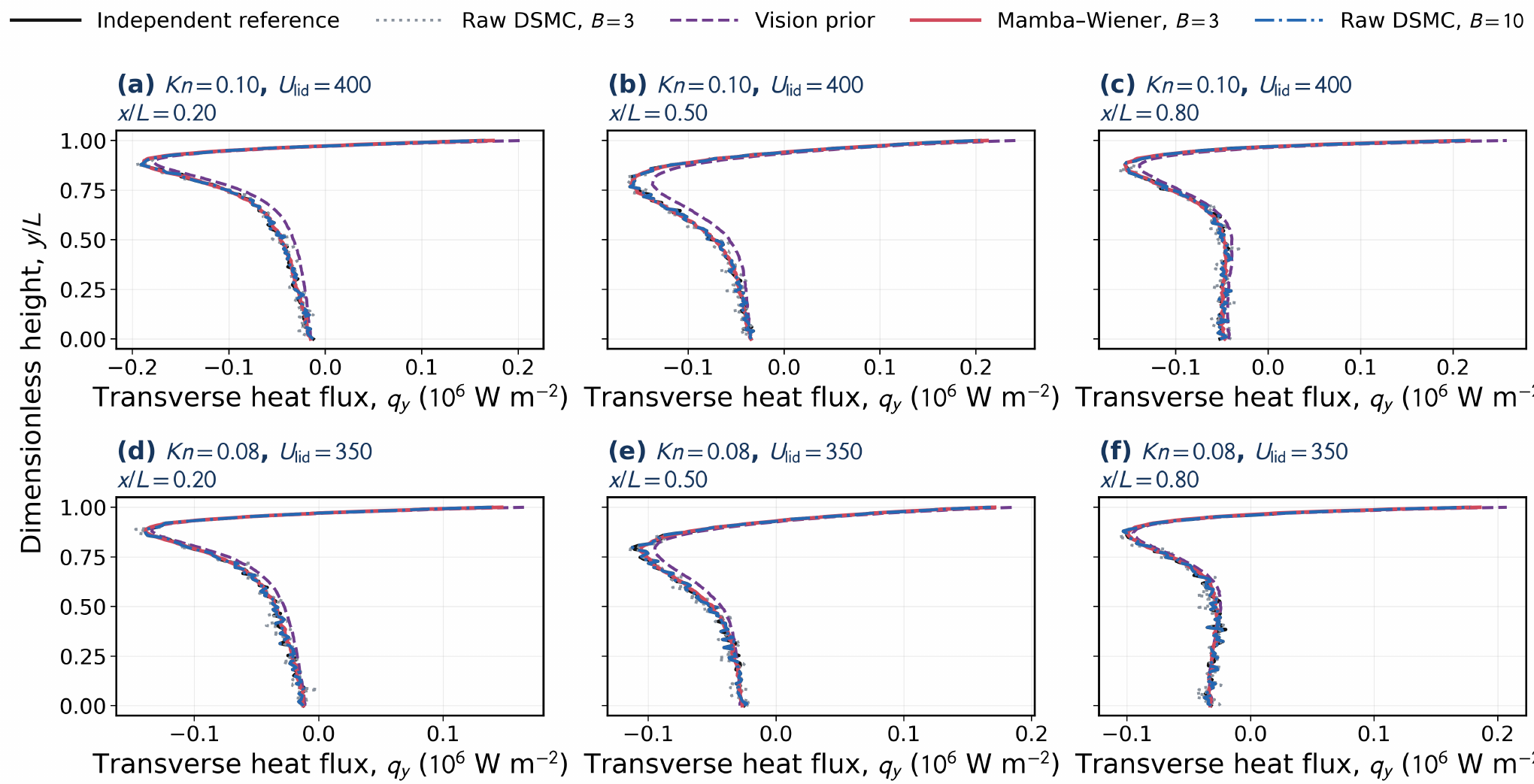}
  \caption{Ensemble-mean $q_y$ along vertical cuts at $x/L=0.20$, 0.50, and 0.80.  The ordinate $y/L$ is dimensionless height.  The upper and lower rows correspond to $(\mathrm{Kn},U_{\mathrm{lid}})=(0.10,400\,\mathrm{m\,s^{-1}})$ and $(0.08,350\,\mathrm{m\,s^{-1}})$, respectively.  Four seeds enter each mean and heat flux is in physical units.}
  \label{fig:cavity-vertical-r14}
\end{journalfigurepage}
\begin{journalfigurepage}
  \journalgraphic{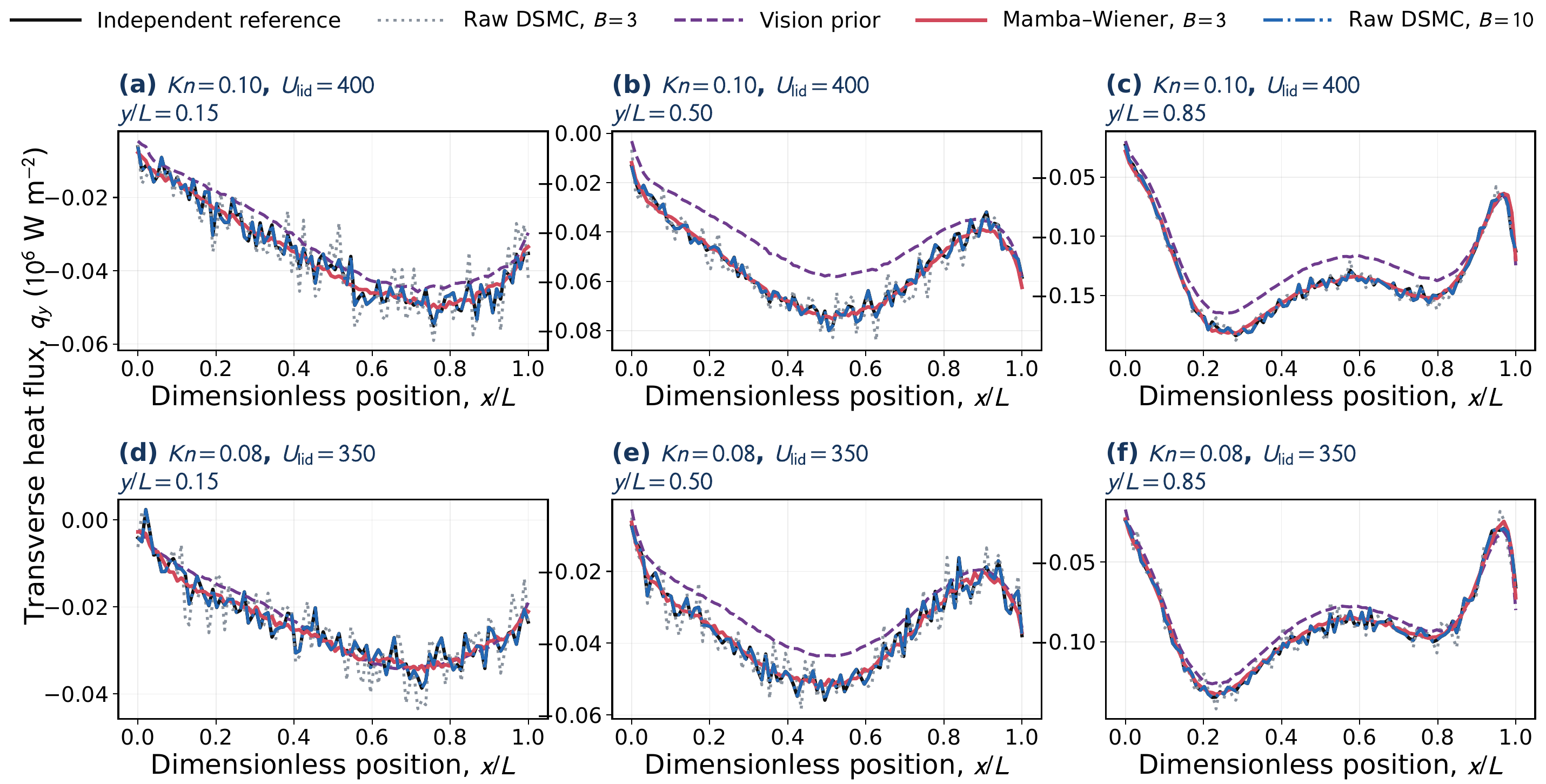}
  \caption{Ensemble-mean $q_y$ along horizontal cuts at $y/L=0.15$, 0.50, and 0.85.  The abscissa $x/L$ is dimensionless position; rows and method definitions are identical to \cref{fig:cavity-vertical-r14}.  Observation conditioning preserves signed lateral redistribution while suppressing \Raw{3} variance.}
  \label{fig:cavity-horizontal-r14}
\end{journalfigurepage}
\begin{journalfigurepage}
  \journalgraphic{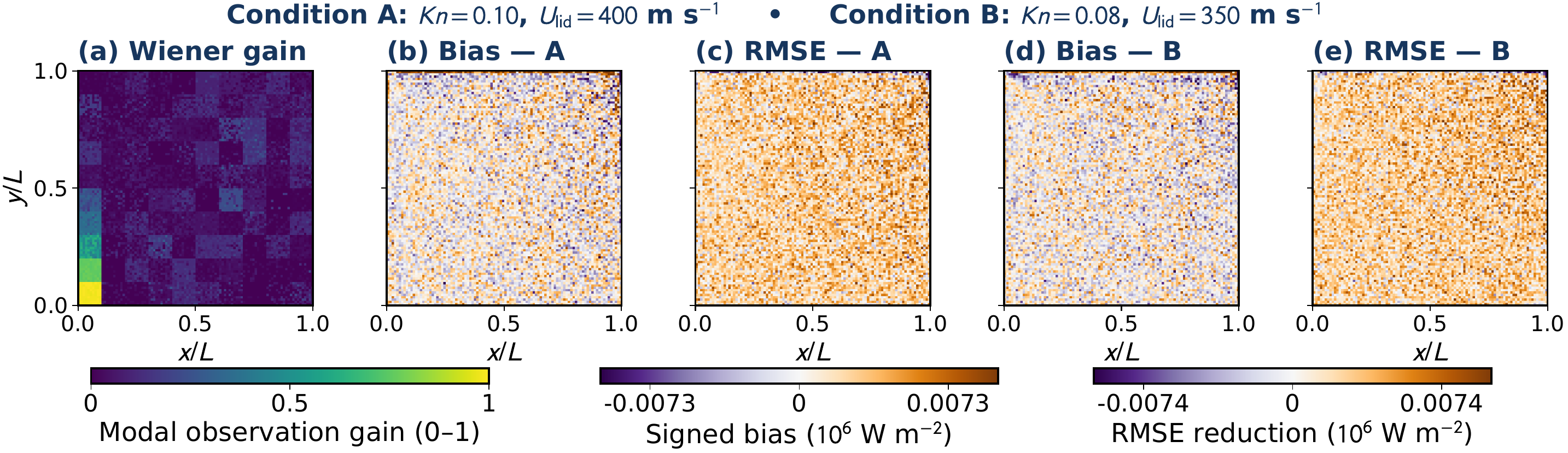}
  \caption{Spectral allocation and cavity error.  The first panel shows the continuous Wiener gain in the two-dimensional discrete cosine transform (DCT) basis; zero rejects and one retains an observed mode.  Remaining panels show signed ensemble bias and local $\Delta\mathrm{RMSE}$ for the two conditions.  Coordinates are normalised by $L$, and positive $\Delta\mathrm{RMSE}$ denotes improvement over \Raw{10}.}
  \label{fig:cavity-spectrum-r14}
\end{journalfigurepage}

\subsection{Mach-10 cylinder reconstruction}
\label{sec:cylinder-results-r14}

Applying the cavity operator directly to the cylinder increased global $q_y$ and near-wall $q_n$ errors to 2.245 and 3.377 times those of \Raw{10}.  A Cartesian cavity basis does not represent a curved wall, bow shock, compression layer, and wake in their natural coordinates.  The cylinder estimator therefore retains the data-consistency principle but replaces the prior, coordinates, quadrature weights, and spectral transfer.  \Cref{fig:cylinder-representation-r14} shows the native areas, cylinder-centred angle, and near-wall selection.  The area panel is the quadrature argument: the adaptive DS2V cells differ substantially in area between the refined near-body region and the coarse outer field, so equal-weight fitting would let the refined region dominate every spectral coefficient and misrepresent the wake.  The angle panel is the alignment argument: in $(r_c/D,\theta)$ the bow shock, compression layer, and wall layer follow coordinate lines and a two-dimensional DCT represents them with few modes, whereas the same structures cut obliquely across a Cartesian basis and spread over many weakly determined coefficients.  No solid cell enters the reconstruction, and the normal/tangential rotation in \cref{eq:rotate-r14} precedes spectral shrinkage.

In the cylinder figures, $q_i^*$, $i\in\{x,y,n,t\}$, denotes heat flux expressed in one common fixed set of scaled plotting units.  The asterisk is a display convention rather than a physical nondimensionalisation; the same scale is used for every panel and method and therefore cancels from all NRMSE ratios.  The representative field in \cref{fig:cylinder-field-r14} contains the bow-shock, shoulder, and wake structure.  The prior supplies most of the coherent pattern, while the bounded current-trajectory correction is small; this division of labour is what the ridge fit in \cref{eq:ridge-r14} learned, because three-block peer residuals are mostly noise, the half-trace regularisation shrinks the fitted singular values well below unity, and the operator therefore transfers only the fraction of the observation that development data could predict.  Read as machine learning, the transfer matrices are deliberately underfitted: a bounded linear map estimated from four trajectories cannot chase a single realisation, which costs little where the prior is accurate and protects the estimate where it is not.  The six-pair mean in \cref{fig:cylinder-ensemble-r14} shows improvement through both the compression layer and the wake, the two regions where refinement and signed transport differ most.  Near-wall profiles and signed errors in \cref{fig:cylinder-qn-r14,fig:cylinder-qn-error-r14} show reduced angular fluctuation in every evaluated pair rather than in a selected seed or sector: the gain appears as variance reduction about the reference profile rather than a change of its shape, consistent with shrinkage that acts on observation-dominated modes.

The selected estimate has a six-pair arithmetic-mean NRMSE of 0.18483 for global $q_y$ and 0.04831 for near-wall $q_n$ (\cref{tab:cylinder-r14}).  The aggregate ratios $\mathcal R_a$ in \cref{eq:aggregate-ratio-r14} are 0.84642 and 0.79336, and all six pairwise ratios are below one for both endpoints.  The geometric means of those six pairwise ratios are 0.84650 and 0.79558, with descriptive log-scale 95\% intervals $[0.83224,0.86101]$ and $[0.72427,0.87391]$.  The exact one-sided sign-test probability is $2^{-6}=0.015625$ for each endpoint; Holm adjustment across the two tests gives 0.03125.

Most of the improvement comes from the frozen prior: relative to it, observation conditioning reduces mean NRMSE by only approximately 0.16\% globally and 0.12\% near the wall.  The correctly phased correction nevertheless outperforms the phase control at both endpoints and preserves the observed mean exactly.  The phase comparison is the informative one for attribution: both corrections carry identical spectral amplitudes, so the margin of the aligned correction in \cref{fig:cylinder-correction-r14} shows that the observation's value lies in the spatial phase of its low modes, their registration with the shock, wall layer, and wake, rather than in amplitude content a permutation could mimic.  The correction's small magnitude agrees with the fitted singular values of 0.0882--0.1630.  Downstream cuts in \cref{fig:cylinder-wake-r14}, the vector magnitude in \cref{fig:cylinder-magnitude-r14}, and the polar field in \cref{fig:cylinder-polar-r14}, where $q_r^*=q_x^*\cos\theta+q_y^*\sin\theta$ coincides with $q_n^*$ for a circle, show that the result is not confined to one Cartesian component: the estimator acts on $(q_n,q_t)$ before rotating back, so the gain is shared by $q_x$, $q_y$, their magnitude, and the polar profile, a property a fixed Cartesian-component estimator would not have near a curved wall.

\begin{table}[htbp]
\centering
\small
\caption{Mach-10 cylinder accuracy against independent references.  Panel A reports means over six pairs, Panel B reports selected-to-\Raw{10} ratios for each pair, and Panel C gives pair-level inference; native cells are not replicates.}
\label{tab:cylinder-r14}
\begin{tabularx}{\textwidth}{@{}Yrrrr@{}}
\toprule
\multicolumn{5}{@{}l}{\textit{Panel A: six-pair mean}}\\
Estimator & $q_y$ NRMSE & $q_y$ ratio & $q_n$ NRMSE & $q_n$ ratio \\
\midrule
\Raw{3} & 0.31788 & 1.45569 & 0.08761 & 1.43889 \\
\Raw{10} & 0.21837 & 1.00000 & 0.06089 & 1.00000 \\
Frozen prior & 0.18513 & 0.84780 & 0.04836 & 0.79428 \\
Phase control & 0.18799 & 0.86088 & 0.04930 & 0.80968 \\
Selected estimate & \textbf{0.18483} & \textbf{0.84642} & \textbf{0.04831} & \textbf{0.79336} \\
\bottomrule
\end{tabularx}
\vspace{0.55em}
\begin{tabular}{@{}lrrrrrr@{}}
\toprule
\multicolumn{7}{@{}l}{\textit{Panel B: ratio by independent pair}}\\
Endpoint & 01 & 02 & 03 & 04 & 05 & 06 \\
\midrule
Global $q_y$ & 0.84604 & 0.85315 & 0.82404 & 0.84048 & 0.86451 & 0.85136 \\
Near-wall $q_n$ & 0.75568 & 0.89933 & 0.71782 & 0.85537 & 0.73901 & 0.82228 \\
\bottomrule
\end{tabular}
\vspace{0.55em}
\begin{tabular}{@{}lcccc@{}}
\toprule
\multicolumn{5}{@{}l}{\textit{Panel C: pair-level summary and inference}}\\
Endpoint & Improved & Geometric mean & Descriptive 95\% interval & Holm $p$ \\
\midrule
Global $q_y$ & 6/6 & 0.84650 & [0.83224, 0.86101] & 0.03125 \\
Near-wall $q_n$ & 6/6 & 0.79558 & [0.72427, 0.87391] & 0.03125 \\
\bottomrule
\end{tabular}
\end{table}

\begin{journalfigurepage}
  \journalgraphic{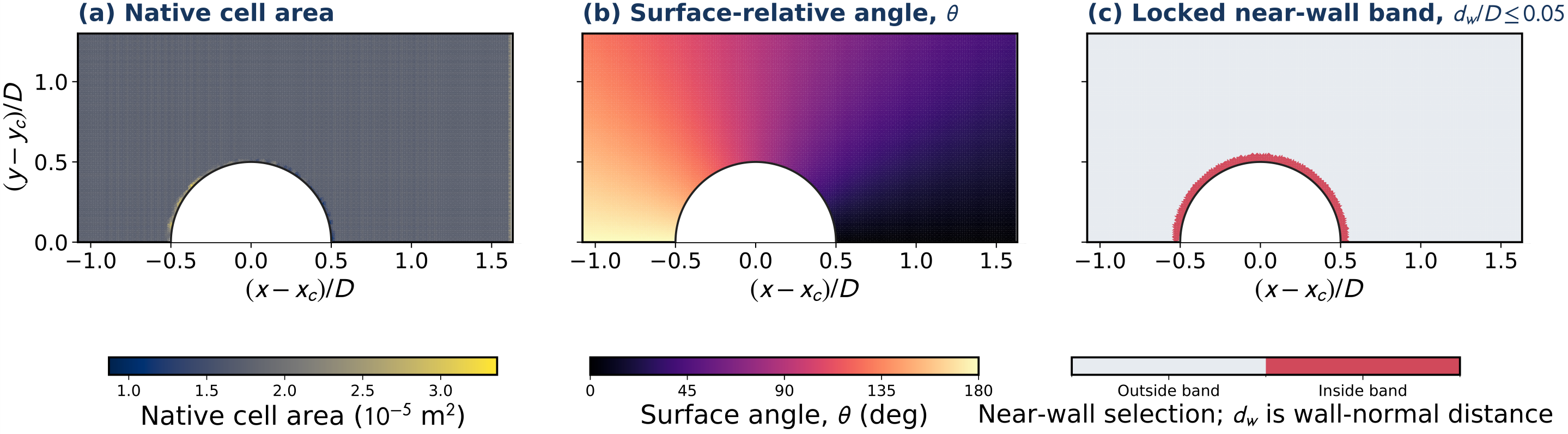}
  \caption{Cylinder-native representation.  (a) Native DS2V fluid-cell centres and areas.  (b) Polar coordinates relative to the cylinder centre, with dimensionless position and surface angle $\theta$.  (c) Near-wall band within $0.05D$ of the surface, where $D$ is cylinder diameter.  Solid cells are excluded and interpolated polar fields use a two-dimensional DCT basis.}
  \label{fig:cylinder-representation-r14}
\end{journalfigurepage}
\begin{journalfigurepage}
  \journalgraphic{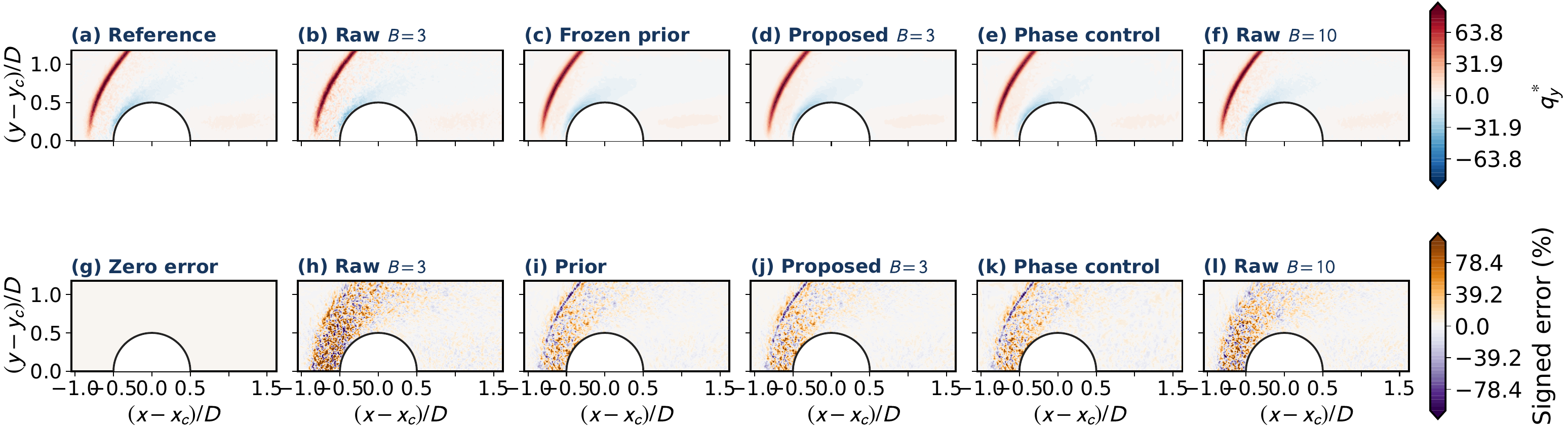}
  \caption{Cylinder transverse heat flux for pair 01.  The upper row shows the independent reference, \Raw{3}, frozen prior, selected three-block estimate, phase-scrambled control, and paired \Raw{10}.  The lower row is signed error normalised by the reference root-mean-square magnitude.  Coordinates are centred on the cylinder and normalised by $D$; solid cells are masked.}
  \label{fig:cylinder-field-r14}
\end{journalfigurepage}
\begin{journalfigurepage}
  \journalgraphic{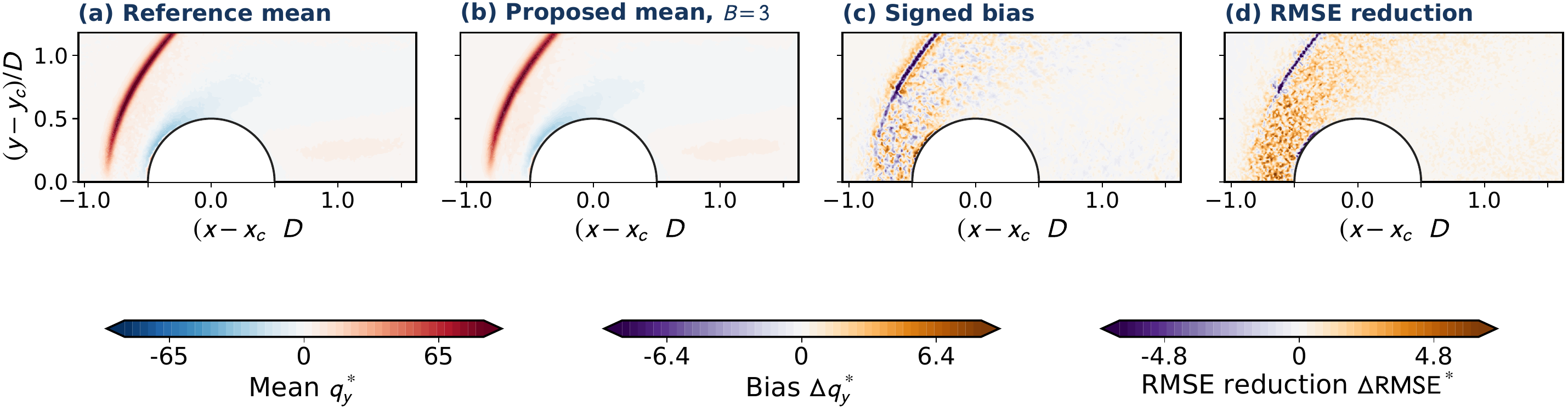}
  \caption{Six-pair cylinder diagnostics for $q_y^*$.  Panels show independent-reference mean, selected three-block mean, signed bias, and local $\Delta\mathrm{RMSE}=\mathrm{RMSE}_{B=10}-\mathrm{RMSE}_{\mathrm{rec}}$.  Positive values denote improvement; coordinates are centred on the cylinder and normalised by $D$.}
  \label{fig:cylinder-ensemble-r14}
\end{journalfigurepage}
\begin{journalfigurepage}
  \journalgraphic{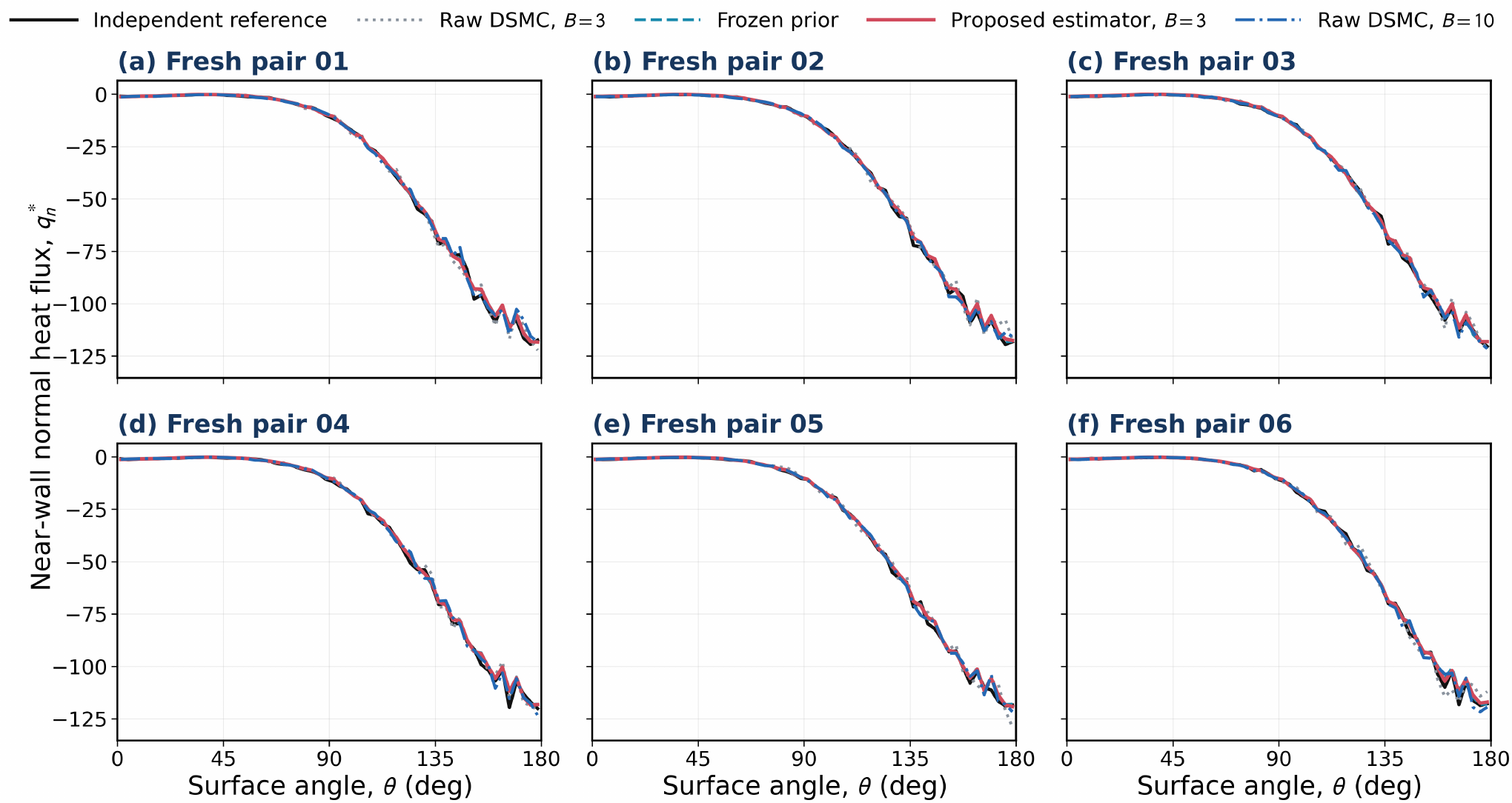}
  \caption{Near-wall $q_n^*(\theta)=\boldsymbol q^*\!\cdot\!\boldsymbol n$ for six cylinder pairs, where $\boldsymbol q^*=(q_x^*,q_y^*)$ and $\boldsymbol n$ is the outward normal.  Panels compare the independent reference, \Raw{3}, frozen prior, selected estimate, and \Raw{10}.  This cell-centred moment differs from the wall-collision tally $q_w$.}
  \label{fig:cylinder-qn-r14}
\end{journalfigurepage}
\begin{journalfigurepage}
  \journalgraphic{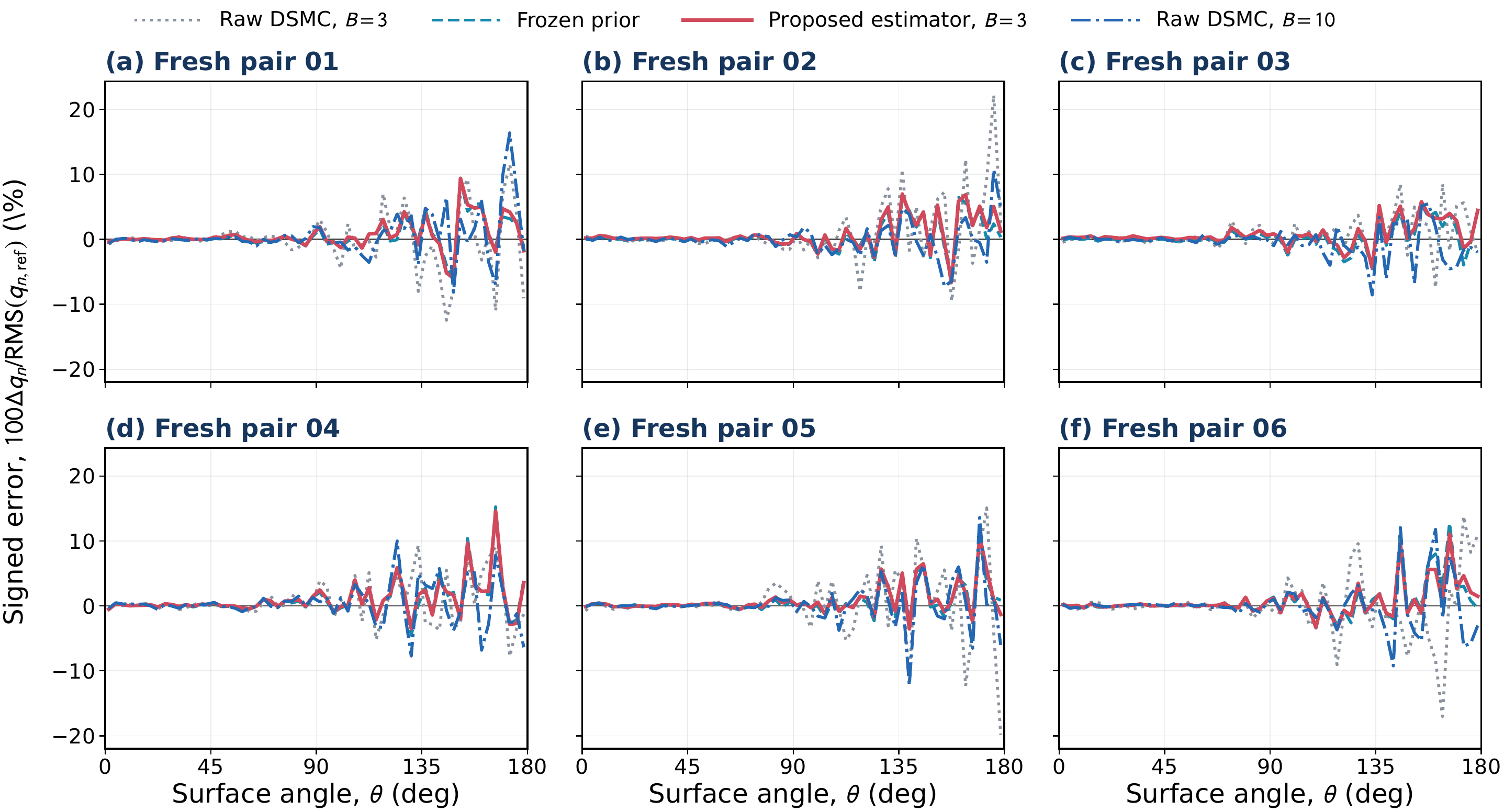}
  \caption{Signed near-wall error $100(q_n-q_{n,\mathrm{ref}})/\mathrm{RMS}(q_{n,\mathrm{ref}})$ for the six cylinder pairs.  Here $q_{n,\mathrm{ref}}$ is the independent-reference profile and RMS denotes its root-mean-square magnitude over $\theta$.  Common method definitions and limits permit direct pairwise comparison.}
  \label{fig:cylinder-qn-error-r14}
\end{journalfigurepage}
\begin{journalfigurepage}
  \journalgraphic{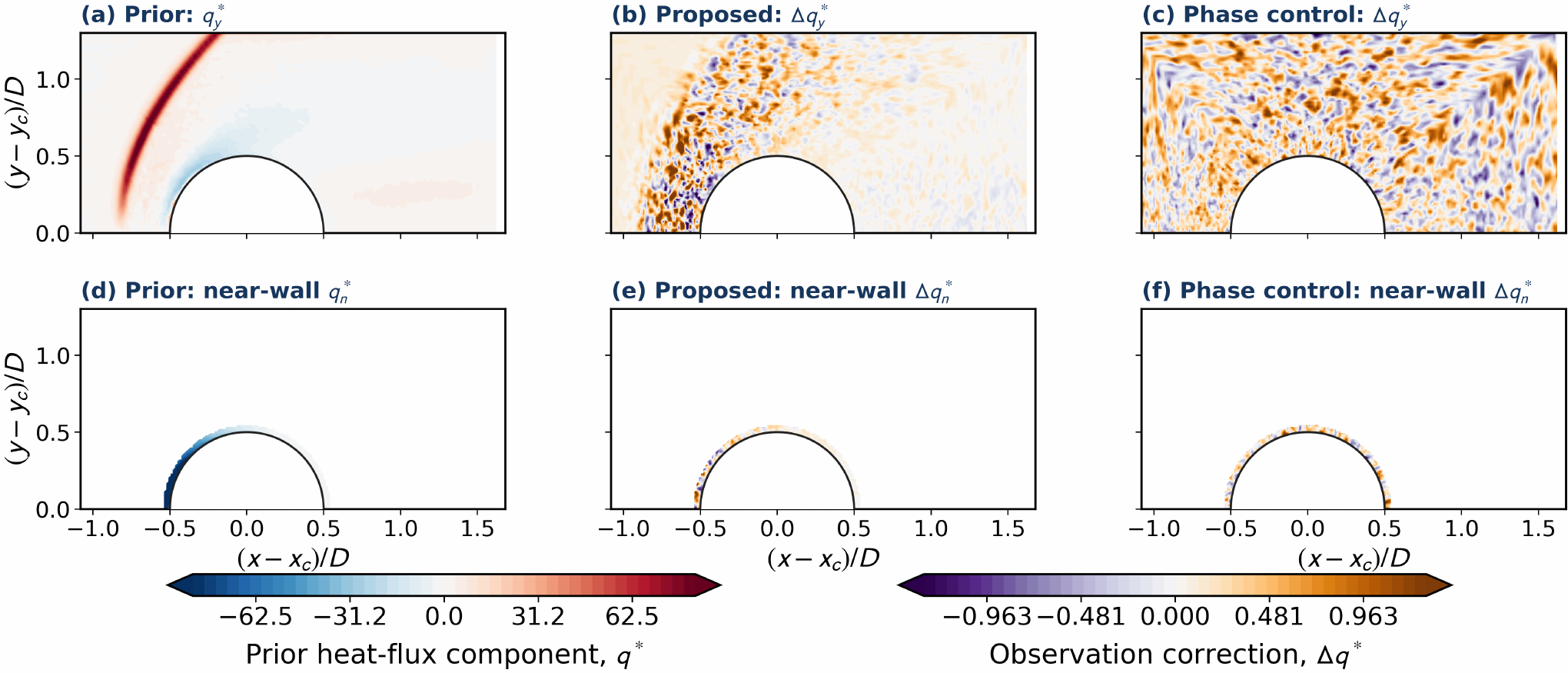}
  \caption{Prior and current-observation correction for cylinder pair 01.  The upper row shows prior $q_y^*$, selected-minus-prior $\Delta q_y^*$, and the phase-control correction.  The lower row shows the corresponding near-wall $q_n^*$ quantities.  Separate matched scales distinguish the repeatable prior from the bounded correction.}
  \label{fig:cylinder-correction-r14}
\end{journalfigurepage}
\begin{journalfigurepage}
  \journalgraphic{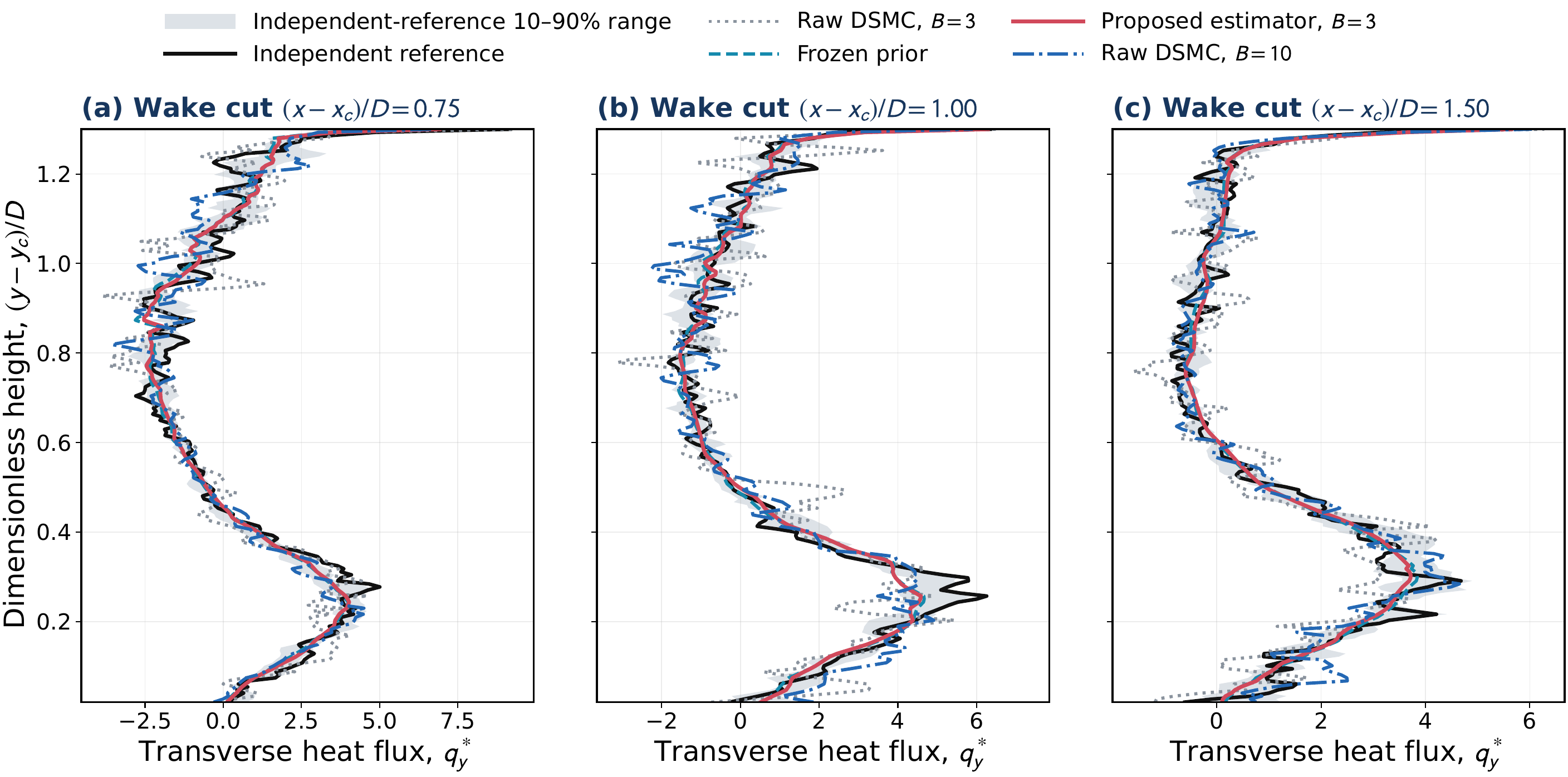}
  \caption{Transverse heat flux $q_y^*$ along downstream cuts at $(x-x_c)/D=0.75$, 1.00, and 1.50.  The ordinate $(y-y_c)/D$ is dimensionless height.  The shaded band spans the 10th--90th percentiles across six independent reference trajectories; method definitions are common to all panels.}
  \label{fig:cylinder-wake-r14}
\end{journalfigurepage}
\begin{journalfigurepage}
  \journalgraphic{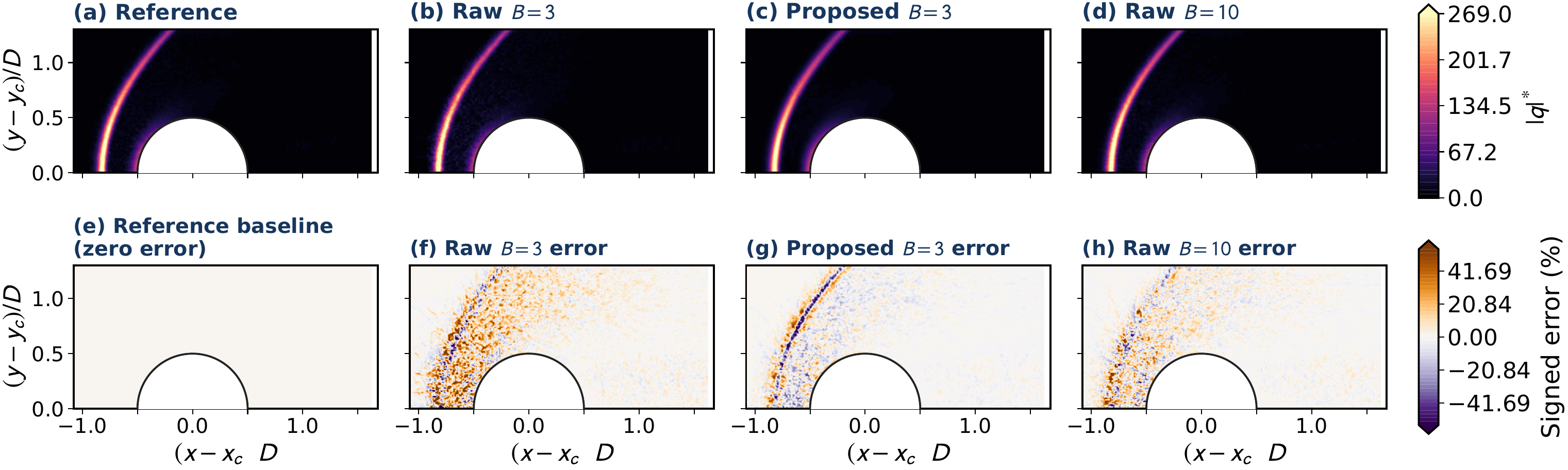}
  \caption{Heat-flux magnitude $|\boldsymbol q^*|=[(q_x^*)^2+(q_y^*)^2]^{1/2}$ and signed error for pair 01 on the native grid.  The upper row compares the independent reference, \Raw{3}, selected estimate, and \Raw{10}; the lower row is normalised by the reference root-mean-square magnitude.}
  \label{fig:cylinder-magnitude-r14}
\end{journalfigurepage}
\begin{journalfigurepage}
  \journalgraphic{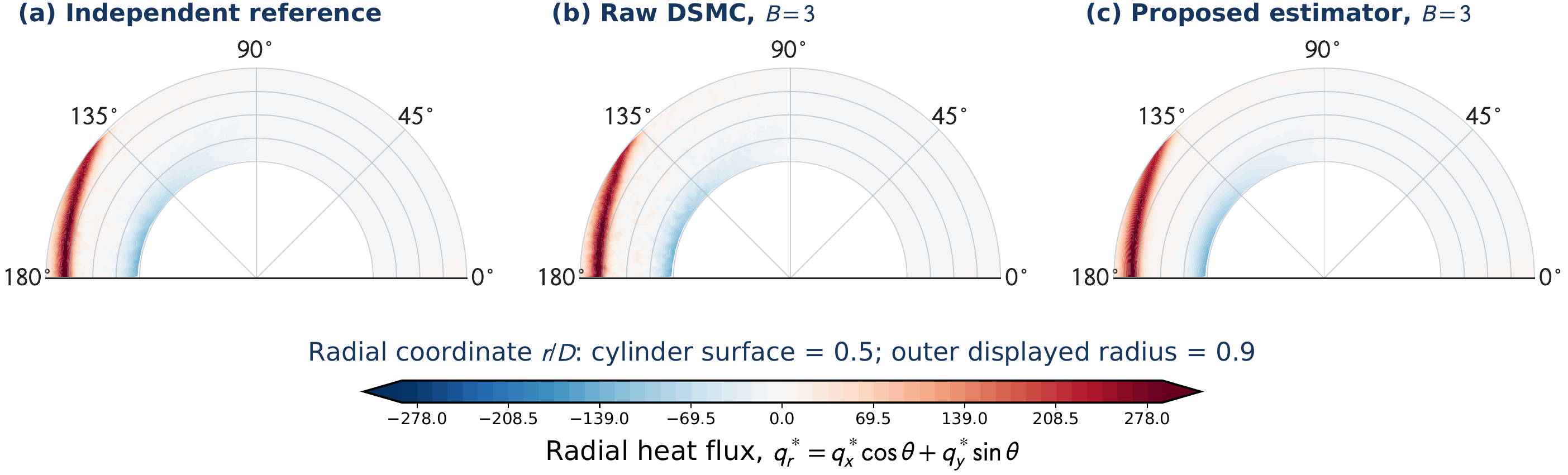}
  \caption{Normal heat flux in native polar coordinates.  The radial component $q_r^*=q_x^*\cos\theta+q_y^*\sin\theta$ equals $q_n^*$ for the outward normal.  Panels compare the independent reference, \Raw{3}, and selected estimate; radius is normalised by $D$ and $\theta=180^{\circ}$ is upstream.}
  \label{fig:cylinder-polar-r14}
\end{journalfigurepage}

\subsection{Support monitoring and the Mach-12 hierarchy}
\label{sec:mach12-results-r14}

Development scores from $N_{\mathrm{dev}}=8$ Mach-8 and Mach-10 trajectories give the support threshold $s_\star=0.435185$.  All six held-out Mach-10 observations have $s_j=0$ and are accepted, with $R_j^K$ from \cref{eq:gain-envelope-ratio-r14} between 0.45 and 0.77.  All four Mach-12 observations have $s_j=1$: each of their 27 monitored component--zone gains is outside the development envelope and $R_j^K$ lies between 8.14 and 8.31 (\cref{fig:support-r14}).  References are absent from both decisions.  Independent scoring afterwards shows that abstention is not numerical failure: relative to \Raw{3}, the outside-support estimate reduces NRMSE by 5.8\% for the nine-field geometric mean, 15.9\% for near-wall $q_n$, and 13.2\% for global $q_y$.  The monitor withholds an in-support interpretation rather than discarding the estimate.

With separate same-condition Mach-12 references, the prior and corrected estimate reduce error for every retained field and all 12 evaluation trajectories.  Ratios in \cref{tab:mach12-r14} range from 0.461 to 0.501, with nine-field geometric means of 0.479 and 0.481.  The current-observation correction is marginal: the prior is slightly better for most fields, whereas the correction is slightly better for $P_{xy}$, $q_x$, and near-wall $q_n$.  The supported result is therefore an approximately twofold error reduction from same-condition information, not systematic superiority of a small residual correction.

Temporal dependence in the reference pool is weak for volume moments but strong for the surface tally.  Across the four 40-block Mach-12 reference trajectories, the pooled fields in \cref{tab:mach12-beff-r14} retain $B_{\mathrm{eff}}=134$--160 effective blocks among 160 nominal blocks, whereas the native wall-collision tally has $B_{\mathrm{eff}}=12.8$.  Pooled volume-moment autocorrelations fall to zero within one to two block lags, whereas the wall tally decorrelates slowly.  The paired ratios are reported as directly observed; no model-based correction for reference noise is applied.

The wall tally $q_w$ differs from nearest-cell $q_n$.  For either angular profile, the upper-surface mean is $\langle q\rangle_\theta=\pi^{-1}\int_0^\pi q(\theta)\,\mathrm d\theta$, with $\theta$ in radians.  The symmetry-constrained treatment in \cref{eq:stagnation-r14} changes the upper-surface mean by only $-0.009\%$.  The resulting means are $\langle q_w\rangle_\theta=21.6\,\mathrm{kW\,m^{-2}}$ and $\langle-q_n\rangle_\theta=17.3\,\mathrm{kW\,m^{-2}}$ (\cref{fig:mach12-wall-r14}).

\begin{table}[htbp]
\centering
\small
\caption{Mach-12 geometric-mean NRMSE ratios to \Raw{10} across 12 independent evaluation trajectories with same-condition information.  The prior contains no current three-block observation; the corrected estimate adds the bounded residual in \cref{eq:same-condition-r14}.  Lower values indicate smaller error against the independent 80-block target.}
\label{tab:mach12-r14}
\begin{tabularx}{\textwidth}{@{}Ycc@{}}
\toprule
Field & Same-condition prior & Corrected \Raw{3} estimate \\
\midrule
Number density $n$ & \textbf{0.49606} & 0.50076 \\
Streamwise velocity $u$ & \textbf{0.47386} & 0.47934 \\
Transverse velocity $v$ & \textbf{0.46991} & 0.47384 \\
Translational temperature $T$ & \textbf{0.49142} & 0.49354 \\
Streamwise normal stress $P_{xx}$ & \textbf{0.48573} & 0.48591 \\
Shear stress $P_{xy}$ & 0.47246 & \textbf{0.47185} \\
Transverse normal stress $P_{yy}$ & \textbf{0.47860} & 0.48193 \\
Streamwise heat flux $q_x$ & 0.48296 & \textbf{0.48133} \\
Transverse heat flux $q_y$ & \textbf{0.46127} & 0.46344 \\
\midrule
Geometric mean, nine fields & \textbf{0.47903} & 0.48121 \\
Near-wall normal heat flux $q_n$ & 0.46250 & \textbf{0.46163} \\
\bottomrule
\end{tabularx}
\end{table}

\begin{table}[htbp]
\centering
\small
\caption{Effective number of independent sampling blocks $B_{\mathrm{eff}}=B/\tau_{\mathrm{int}}$ from \cref{eq:beff-r14}, pooled across the four 40-block Mach-12 reference trajectories with $B=160$ nominal blocks.  Volume moments and near-wall $q_n$ are cell-centred fields; the wall-collision tally $q_w$ is the surface accumulator.}
\label{tab:mach12-beff-r14}
\begin{tabularx}{\textwidth}{@{}Yc@{}}
\toprule
Field & $B_{\mathrm{eff}}$ of 160 \\
\midrule
Number density $n$ & 137.0 \\
Streamwise velocity $u$ & 133.7 \\
Transverse velocity $v$ & 148.3 \\
Translational temperature $T$ & 138.4 \\
Streamwise normal stress $P_{xx}$ & 148.3 \\
Shear stress $P_{xy}$ & 157.3 \\
Transverse normal stress $P_{yy}$ & 157.6 \\
Streamwise heat flux $q_x$ & 153.0 \\
Transverse heat flux $q_y$ & 160.0 \\
Near-wall normal heat flux $q_n$ & 160.0 \\
\midrule
Wall-collision heat flux $q_w$ & 12.8 \\
\bottomrule
\end{tabularx}
\end{table}

\begin{journalfigurepage}
  \journalgraphic{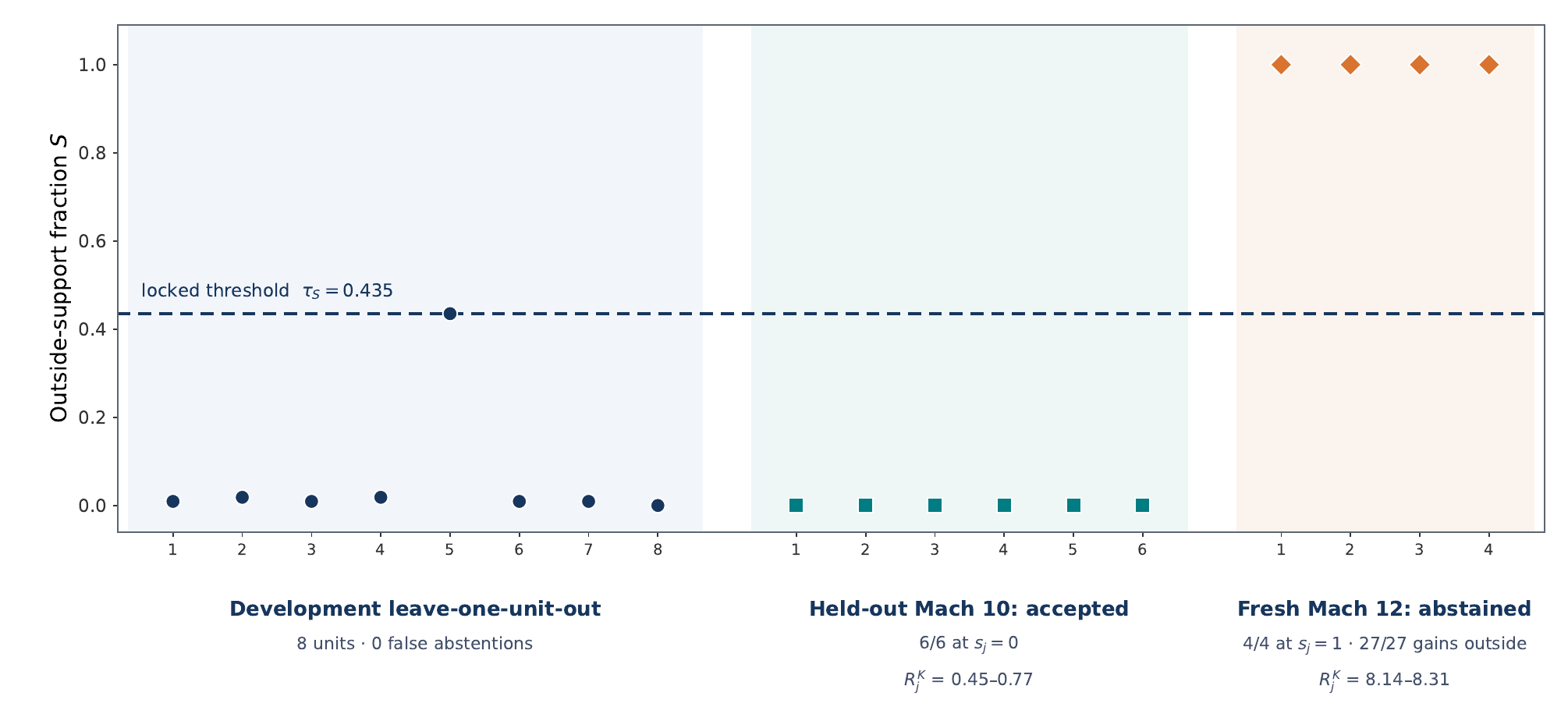}
  \caption{Reference-independent support validation.  Development-only calibration fixes $s_\star=0.435185$.  The monitor accepts all six held-out Mach-10 observations at $s_j=0$; their maximum gain-to-envelope ratios $R_j^K$ are 0.45--0.77.  All four independently generated Mach-12 observations have $s_j=1$ and are classified outside support before reference scoring.}
  \label{fig:support-r14}
\end{journalfigurepage}
\begin{journalfigurepage}
  \journalgraphic{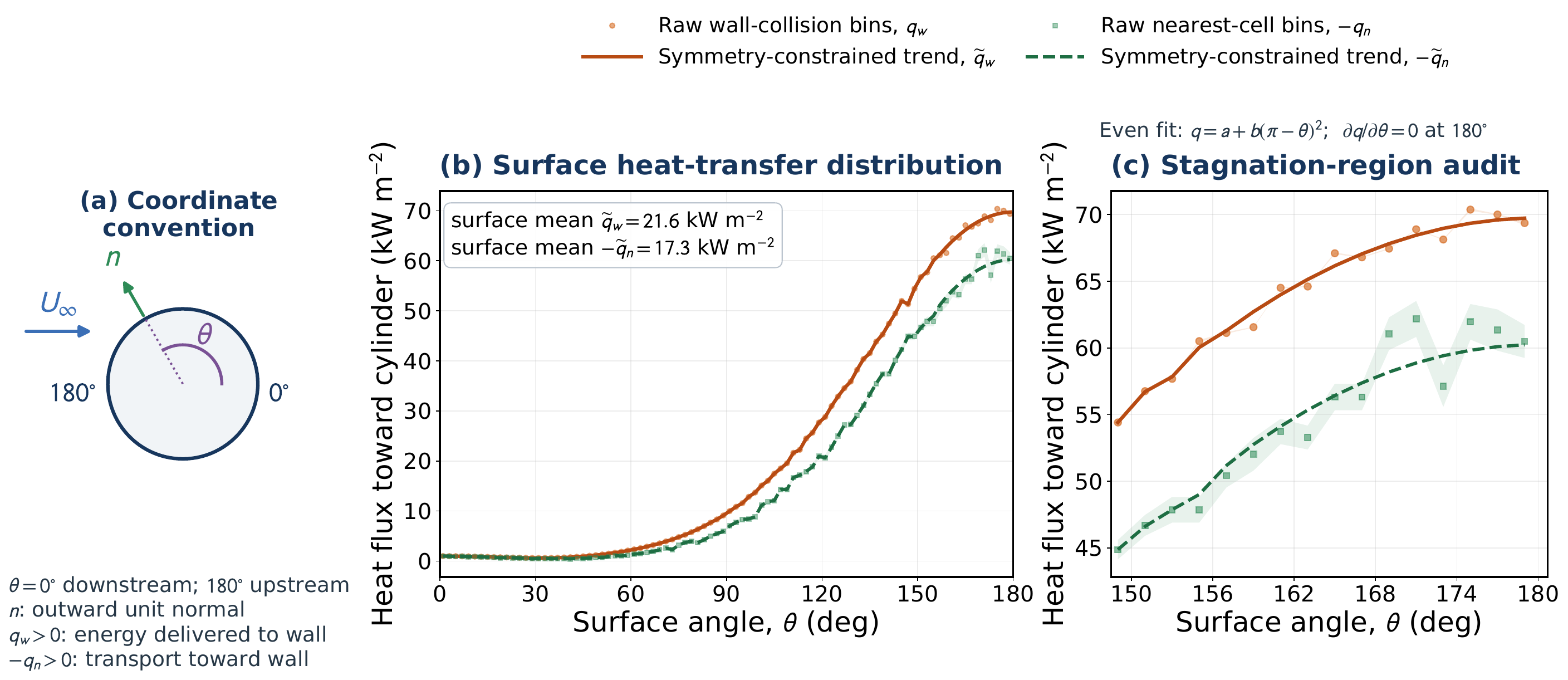}
  \caption{Mach-12 upper-surface heat transfer.  (a) Conventions: $\boldsymbol n$ is outward, $\theta=180^{\circ}$ is upstream, $-q_n>0$ denotes cell-centred transport toward the wall, and $q_w>0$ denotes energy delivered to the wall.  (b) Raw bins, symmetry-constrained trends, and 95\% sampling bands $\overline q(\theta)\pm1.96s_q(\theta)/\sqrt{B_{\mathrm{eff}}}$ from the held-out 80-block pool; $B_{\mathrm{eff}}=6.5198$ for $q_w$ and 54.6327 for the angular $-q_n$ profile.  (c) Stagnation enlargement with \cref{eq:stagnation-r14}; raw values remain visible.}
  \label{fig:mach12-wall-r14}
\end{journalfigurepage}

\FloatBarrier

\section{Discussion}
\label{sec:discussion}

\subsection{Geometry, data support, and the origin of reconstruction gains}

The results support geometry-native regularisation rather than a geometry-independent image filter.  In the cavity, Cartesian modes provide a natural separation between repeatable low-frequency structure and sampling-dominated content.  Around the cylinder, the same physical idea is effective only after the heat-flux vector is expressed in normal and tangential components, solid cells are excluded, and native cell areas are retained.  The distinction matters near a curved wall and through the bow shock, where a fixed Cartesian component mixes radial and angular structure.  The transferable element of the method is therefore the prior-plus-observation construction in \cref{eq:unified-r14}, together with bounded residual gains and exact treatment of observed additive modes.  The prior, coordinates, masks, and transfer operator remain specific to the geometry on which they were constructed.

The Mach-10 comparison also clarifies where the accuracy originates.  The prior-only and observation-conditioned ratios are nearly identical, so most of the reduction relative to the ten-block direct comparator comes from repeatable structure learned from the development trajectories.  The three-block correction is small, although it is not immaterial: it restores the measured Cartesian mean exactly, and a phase-scrambled correction is consistently worse than the spatially aligned correction.  These controls show that the present observation carries usable information; they do not imply that it creates the dominant flow structure.  They also expose a limitation.  Because the fitted nonzero-mode gains are small, a short observation cannot be expected to repair an arbitrarily biased prior after a substantial operating-condition change.

The support score addresses that limitation by separating reconstruction from domain compatibility.  All six held-out Mach-10 observations are consistent with the Mach-8/Mach-10 gain envelope, whereas all four Mach-12 observations lie outside it in every monitored field--zone combination.  This separation is stronger than a marginal threshold crossing, but it should not be interpreted as an error estimator.  Independent scoring shows that the abstained estimate still improves all three reported Mach-12 summaries relative to the raw three-block observation.  Abstention therefore withholds the in-support accuracy claim; it does not declare the reconstructed field useless.  Acceptance, in turn, indicates similarity to the calibrated gain regime and guarantees no prescribed error for an individual trajectory.

The present support result is deliberately narrow.  It demonstrates supported-condition retention at Mach 10 and detection of the tested shift to Mach 12 for the same gas, Knudsen number, wall model, and cylinder geometry.  Eight Mach-8/Mach-10 development trajectories are insufficient to establish a universal boundary in Mach number, and no conclusion follows automatically for a different Knudsen number, wall temperature, gas species, body shape, or transient regime.  Extension to any of those settings requires new development coverage and an independent evaluation of both supported observations and relevant shifts.  The separate same-condition Mach-12 analysis illustrates the complementary case in which in-condition information is explicitly available: all nine retained fields attain ratios near one-half, although the prior itself supplies nearly all of that gain.  This reinforces the central role of data support without undermining the case for observation consistency.

\subsection{Statistical interpretation and limitations}

The cavity results identify a useful mechanism but remain developmental.  The spectral gain allocation was selected within the cavity analysis, and only four seeds are available at each condition; even improvement for every seed cannot yield a one-sided exact probability below 0.0625.  In addition, two trajectories did not satisfy the conservative temperature-extremum stationarity diagnostic noted in \cref{sec:design}.  These two failures do not by themselves show that the reported $q_y$ comparisons are spurious, but they prevent the cavity study from serving as an unqualified independent confirmation.  Its strongest conclusions are that a vision prior can retain coherent structure while miscalibrating amplitude and offset, and that observation-supported low modes can correct this failure more effectively than smoothness or a scalar energy residual alone.

The Mach-10 evidence is stronger because it uses six disjoint observation/reference pairs and reports every pair.  Both co-primary endpoints improve in all six pairs, giving Holm-adjusted one-sided sign-test probabilities of 0.03125.  Nevertheless, the sample remains small, and the experiment supports only the declared finite-time state.  The evaluated outputs span $11.268\le tU_\infty/D\le11.390$; they do not demonstrate asymptotic stationarity or behaviour at a later target such as $tU_\infty/D=30$.  Treating native cells as independent replicates would not remove this limitation because all cells in a field share the same stochastic trajectory, prior, and reconstruction operator.  Additional independent pairs and later-time sampling are needed to quantify how stable the effect size is beyond the present window.

All reported errors are measured against finite stochastic references rather than an exact Boltzmann solution.  Independent references prevent their particle noise from entering prediction, and using the same reference for an estimator and its comparator makes each ratio a fair paired comparison.  Reference variance nevertheless contributes to both observed errors and can attenuate their separation.  The cavity leave-one-seed-out reference and the independent cylinder references also have different variance structures, so their absolute NRMSE values should not be compared as if they had a common truth budget.  The directly observed paired ratios remain the reproducible quantities; any correction for reference noise requires additional assumptions and is appropriately secondary.

The use of three rather than ten sampling blocks is a 70\% reduction in the declared averaging-block budget.  It is not a demonstrated 70\% reduction in end-to-end wall-clock cost.  Particle motion, collision sampling, transient evolution, input/output, and scheduler overhead do not scale linearly with the retained averaging window, while prior construction has an offline cost that must be amortised over later evaluations.  A wall-clock speedup claim would require matched runs with identical initialisation and stopping criteria, together with measured simulation and reconstruction times.  The present computational claim is therefore restricted to the sampling budget and the accompanying error comparison.

Finally, cell-centred normal heat flux and wall heat transfer must not be conflated.  The quantity $q_n=\boldsymbol q\cdot\boldsymbol n$ is a full-range velocity moment evaluated at the nearest fluid-cell centres, whereas $q_w$ is a half-range molecular energy-exchange statistic accumulated at wall collisions.  Their different locations and estimators explain why $-q_n$ need not equal $q_w$.  Temporal dependence sharpens the contrast: the four-trajectory Mach-12 reference pool contains 160 nominal blocks, of which the volume moments retain 134--160 effectively independent blocks but the wall tally retains only 12.8.  Surface heating therefore needs its own collision tally, its own autocorrelation analysis, and its own sign convention and uncertainty estimate.  The symmetry-constrained stagnation treatment changes the surface mean by less than 0.01\% and leaves the raw angular bins visible, but it does not convert the nearest-cell moment into a wall boundary measurement.

\section{Conclusions}
\label{sec:conclusions}

This work introduced a non-intrusive, geometry-native machine learning reconstruction framework for the complete retained DSMC moment hierarchy.  The estimator combines structure learned from independent trajectories with a bounded residual from the current low-budget observation, while preserving directly observed zero-frequency content.  Forming central moments only after additive accumulators are combined maintains consistency among density, velocity, temperature, pressure, and heat flux.  The formulation also makes explicit why energy conservation alone cannot identify an individual heat-flux component: divergence-free errors are invisible to the scalar energy balance.

The cavity study showed that machine-learned image restoration alone can recover spatial structure yet carry appreciable amplitude and offset bias.  With the observation-conditioned spectral correction added, the transverse-heat-flux NRMSE fell to 0.658 and 0.672 times the ten-block value at the two conditions considered.  For the circular cylinder, reformulating the estimator in normal and tangential coordinates produced a stronger independent test: the three-block estimate improved both global transverse heat flux and near-wall normal heat flux in all six Mach-10 observation/reference pairs.  The arithmetic-mean NRMSE ratios were 0.846 and 0.793, and the Holm-adjusted one-sided exact probability was 0.03125 for each endpoint.

The support monitor complemented reconstruction accuracy with a deployment decision that did not use a reference.  It accepted all six held-out Mach-10 observations and abstained on all four fresh Mach-12 observations before their references were accessed, demonstrating separation between supported and shifted operating conditions under the fixed rule.  Post-decision scoring nevertheless showed that an abstained estimate can remain numerically useful, whereas the separate same-condition Mach-12 analysis reduced the error of all nine retained fields to roughly half the ten-block value.  Support and accuracy are therefore related but not interchangeable, and they should be reported separately.

The demonstrated reduction from ten to three sampling blocks concerns the averaging budget, not total solver wall time, and the conclusions remain restricted to the geometries, conditions, and finite-time windows examined here.  Extension to other configurations requires geometry-appropriate coordinates, newly calibrated condition-specific statistics, and independent evaluation.  Surface heating also requires its native wall-collision tally and temporal-correlation analysis rather than substitution by the nearest cell-centred moment.  Within these limits, the results show that a single computational framework can reconstruct the complete retained DSMC field vector from a reduced sampling budget, keep its moments kinetically consistent, withstand independent validation, and report, from the observation alone, whether an estimate lies inside its validated domain.

\enlargethispage{2\baselineskip}
\section*{Data and code availability}

Analysis code and compact data required to reproduce the figures and metrics are available at \url{https://github.com/Ehsan-Roohi/DSMC_Python}; raw DSMC checkpoints are available from the author on reasonable request.

\clearpage
\setlength{\bibsep}{0.35\baselineskip}
\bibliographystyle{unsrtnat}
\bibliography{references}

\end{document}